\documentclass{aa}
\usepackage{graphicx}

\newcommand{\be}{\begin{equation}}
\newcommand{\ee}{\end{equation}}
\newcommand{\beq}{\begin{eqnarray}}
\newcommand{\eeq}{\end{eqnarray}}
\newcommand{\bez}{\begin{eqnarray*}}
\newcommand{\eez}{\end{eqnarray*}}
\newcommand{\bc}{\begin{center}}
\newcommand{\ec}{\end{center}}
\newcommand{\ovl}{\overline}
\newcommand{\unl}{\underline}
\newcommand{\Sp}{\mathop{\rm Sp}\nolimits}

\def\intl{\int\limits}
\def\disp{\displaystyle}
\def\Dr#1#2{\frac{\partial#1}{\partial#2}}
\def\DR#1#2{\frac{\d#1}{\d#2}}

\def\vnb{\mbox{\boldmath $\nabla$}}
\def\ve{\mbox{\boldmath $e$}}
\def\vk{\mbox{\boldmath $k$}}
\def\vkpr{\mbox{\boldmath $k$}'}
\def\vr{\mbox{\boldmath $r$}}
\def\vrpr{\mbox{\boldmath $r$}'}
\def\vp{\mbox{\boldmath $p$}}
\def\vppr{\mbox{\boldmath $p$}'}
\def\vv{\mbox{\boldmath $v$}} 

\def\va{\mbox{\boldmath $a$}}
\def\vgamma{\mbox{\boldmath $\gamma$}}

\def\vrs{\mbox{\boldmath $\scriptstyle r$}}
\def\vps{\mbox{\boldmath $\scriptstyle p$}}
\def\vvs{\mbox{\boldmath $\scriptstyle v$}} 
 
\def\stn{{\bf n}} 

\def\diag{{\rm diag}}
\def\Ne{{N_-}}
\def\d{{\rm d}}
\def\e{{\rm e}}

\def\cV{{\cal V}}

\def\cS{{\cal S}}
\def\cM{{\cal M}}
\def\cN{{\cal N}}
\def\nx{n_{\rm x}}
\def\ny{n_{\rm y}}
\def\nz{n_{\rm z}}
\def\Lx{L_{\rm x}}
\def\Ly{L_{\rm y}}
\def\Lz{L_{\rm z}}

\def\he{\hat{e}}
\def\hk{\hat{k}}
\def\hp{\hat{p}}

\def\ua{\unl{a}}
\def\ue{\unl{e}}
\def\uk{\unl{k}}
\def\up{\unl{p}}

\def\ur{\unl{r}}

\def\unb{\unl{\nabla}}
\def\ugam{\unl{\gamma}}

\def\Te{T_{\rm e}}
\def\Tb{T_{\rm b}}
\def\kB{k_{\rm B}}
\def\taut{\tau_{\rm T}}
\def\mA{{\bf A}}
\def\mN{{\bf N}}
\def\mF{{\bf F}}
\def\mL{{\bf L}}

\def\amu{{a}_{\mu}}

\def\anu{{a}_{\nu}}

\def\Mo#1#2{{M\raise8.5pt\hbox{\hskip-7pt$\circ$}_{#1}
^{\raise-5pt\hbox{\scriptsize\,#2}}}}

\def\To#1#2{{T\raise8.5pt\hbox{\hskip-6pt$\circ$}_{#1}
^{\raise-5pt\hbox{\scriptsize\,${#2}$}}}}

\def\tm#1#2{{t\raise8.5pt\hbox{\hskip-4pt$\circ$}_{#1}
^{\raise-5pt\hbox{\scriptsize\,#2}}}}
 
\def\matr44#1#2#3#4#5#6{M_{#1}^{#2}\!\biggl({#4 \atop 
#3}\biggl|{#6\atop#5}\biggr) }
\def\tatr44#1#2#3#4#5#6{T_{#1}^{#2}\!\biggl({#4 \atop 
#3}\biggl|{#6\atop#5}\biggr) }

\def\Motr#1#2#3#4#5#6{{ {M\raise7.5pt\hbox{\hskip-6pt$\circ$}_{#1}
^{\raise-5pt\hbox{\scriptsize\,${#2}$}}}\!\biggl({#4 \atop 
#3}\biggl|{#6\atop#5}\biggr) }}

\def\Totr#1#2#3#4#5#6{{ {T\raise7.5pt\hbox{\hskip-6pt$\circ$}_{#1}
^{\raise-5pt\hbox{\scriptsize\,${#2}$}}}\!\biggl({#4 \atop 
#3}\biggl|{#6\atop#5}\biggr) }}

\def\rhotp#1#2#3#4{\rho^{#1}_{#2}\!\!\left(\begin{array}{l}#3\\#4\end{array}
\right)}
\def\rhoma#1#2#3#4#5#6#7{#1^{#2}_{#3}\!\!\left({#4\atop#5}\biggl|{#6\atop#7}
\right)}
\def\rhomat#1#2#3#4#5#6#7{\rho^{#1}_{#2}\!\!\left({#3\atop#4}\biggl|{#5\atop#6}
\biggr|#7\right)}
\def\rhokkp#1#2#3#4#5#6#7#8{\rho^{#1}_{#2}\!\!\left({#3\atop#4}{#5\atop#6}
\biggl|{#7\atop#8}\right)}
\def\rhokpp#1#2#3#4#5#6#7#8{\rho^{#1}_{#2}\!\!\left({#3\atop#4}\biggl| 
{#5\atop#6}{#7\atop#8}\right)}
\def\rhofour#1#2#3#4#5#6{\rho^{#1}_{#2}\!\!\left({#3\atop#4}{#5\atop#6}\right)}
\def\rhotwo#1#2#3#4{\rho^{#1}_{#2}\!\!\left({#3\atop#4}\right)}

\def\cMatr46#1#2#3#4#5#6#7#8{\cM_{#1}^{#2}\!\biggl({#4 \atop 
#3}\biggl|{#6\atop#5}\biggl|{#8\atop#7} \biggr) }
\def\cNatr46#1#2#3#4#5#6#7#8{\cN_{#1}^{#2}\!\biggl({#4\atop#3}\biggl|
{#6\atop#5}\biggl|{#8\atop#7}\biggr)}

\begin{document}

\onecolumn

\title{Relativistic Kinetic Equation for Induced Compton Scattering of 
Polarized Radiation}
\author{Dmitrij~I.~Nagirner\inst{1} \and Juri Poutanen\inst{2,3}}
\date{\today} 
\institute{Sobolev Astronomical Institute, St. Petersburg State University,
Staryj Peterhof, 198904 St. Petersburg, Russia\\
\email{dinmax@dn1756.spb.edu}
\and 
Stockholm Observatory, 106 91  Stockholm, Sweden\\
\email{juri@astro.su.se}
\and 
Astronomy Division, P.O. Box 3000, 90014 University of Oulu, 
Finland}

\authorrunning{D.~I.~Nagirner and J.~Poutanen}
\titlerunning{Relativistic Kinetic Equation for Induced Compton Scattering} 
\abstract{The  relativistic  kinetic  equations  describing  time evolution 
and space
dependence  of the density  matrices of  polarized  photons  and  electrons
interacting via Compton  scattering are deduced from the quantum  Liouville
equation.  The induced  scattering  and exclusion  principle are taken into
account.  The   Bogoliubov   method  is  used  in  the  frame  of   quantum
electrodynamics.  The  equation  for  polarized   radiation   scattered  by
unpolarized   electrons  is  considered   as  a  particular   case  and  is
reformulated in terms of the Stokes  parameters.  The  expressions  for the
scattering amplitudes and cross-sections are derived simultaneously.
\keywords{methods: analytical -- radiation mechanisms: general -- plasmas --
polarization -- scattering}}
\maketitle

\section{Introduction}

Compton scattering is an important  physical process in many  astrophysical
systems,  such as  active  galactic  nuclei,  X-ray  binaries,  and  pulsar
magnetospheres   (see  e.g.  reviews  by  Blandford  \&  Scharlemann  1975;
Pozdnyakov,  Sobol', \&  Sunyaev  1983;  Poutanen  1998).  Inverse  Compton
scattering  (i.e.  scattering  of  soft  photons  by hot  electron  gas) is
believed to be the main  mechanism  of the  X-ray/gamma-ray  production  in
accreting  X-ray  binaries  (e.g.  Sunyaev \& Titarchuk  1980;  Poutanen \&
Svensson 1996),  gamma-ray  bright active galactic  nuclei (e.g.  Sikora et
al.  1997), and  possibly  gamma-ray  bursts  (Stern  1999;  Ghisellini  \&
Celotti  1999).  The hard X-rays can  interact  with the cold  material via
classical Compton  scattering losing their energy and producing a cutoff in
the ``Compton  reflected'' spectrum (e.g.  George \& Fabian 1991; Poutanen,
Nagendra,  \&  Svensson  1996).  Induced  (stimulated)  Compton  scattering
becomes very important when the radiation  brightness  temperature is large
$\kB\Tb/mc^2\gg  1/\taut$,  where  $\taut$ is  Thomson  optical  thickness.
Induced  effects can distort  the low  frequency  part of the  radio-source
spectra even when $\taut$ is small  (Sunyaev  1971) and can  influence  the
heating of electrons  near active  galactic  nuclei and pulsars  (Levich \&
Sunyaev 1971).

The process of multiple  scattering  of radiation is described by a kinetic
equation.  This  equation  can be easily  written if we neglect the induced
scattering  and/or the  polarization  (see, e.g.  Nagirner \& Poutanen 1994
and references  therein).  Induced  scattering  leads to the  appearance of
nonlinear  terms  in  the  kinetic  equation.  Since  the  polarization  of
radiation is described by four  parameters, one must write a vector kinetic
equation, i.e.  a system of four equations.  The kinetic equation combining
these two effects  (polarization  and induced  scattering)  was not derived
self-consistently  up to now.  The aim of the  present  paper is to fill up
this gap.

The first kinetic equation for Compton  scattering with induced  scattering
was  written  by  Kompaneets  (1956)  and  is  known  under  his  name.  He
considered  multiple  scattering of homogeneous and isotropic  radiation in
infinite space filled with homogeneous, nondegenerate thermal electron gas.
The  gas  was  assumed  to  be  non-relativistic  ($\kB\Te\ll  mc^2$),  the
radiation  to be  rather  soft  ($h\nu\ll  mc^2$),  and  the  intensity  of
radiation to be a  sufficiently  smooth  function of frequency.  Because of
small changes of photon  frequency in a single  scattering,  the scattering
integral was transformed to a differential  operator by the Fokker---Planck
method.  The equation was  rediscovered  by Weymann  (1966).  More  general
Fokker---Planck  equations were deduced by Cooper (1971) and Barbosa (1982)
for more energetic  electrons and photons, and by Molodtsov  (1994) for the
anisotropic,  degenerate and moving electrons.  For very cold electron gas,
the Fokker---Planck equation was derived by Ross et al.  (1978).

The induced scattering effects strongly depend on the angular  distribution
of the radiation field.  Babuel-Peyrissac  \& Rouvillois (1969) generalized
the Kompaneets equation to nonhomogeneous and anisotropic radiation.  Their
equation  contains  an  integral  operator  in  angular  variables,  but  a
differential operator in frequency.  A simplified form of that equation was
used  by  a  number  of  authors  to  study   induced   effects  in  pulsar
magnetospheres  and radio sources in active galactic nuclei (Wilson \& Rees
1978; Coppi, Blandford, \& Rees 1993; Sincell \& Krolik 1994; Lyubarskii \&
Petrova 1996; Sincell \& Coppi 1996).

As it is mentioned above, the induced scattering makes equations nonlinear.
This effect can be simply  taken into  account in the  equation if we limit
ourselves to one  parameter,  intensity  $I$,  ignoring  polarization.  The
scattering  rate is  increased  by  stimulated  scattering  by a factor $1+
h^2I(\vk)/2ck^3$  (where  $\vk$ is the photon  momentum  after  scattering)
comparing with the rate when only spontaneous scattering is considered.  It
is this factor that the authors of the mentioned  works  (Kompaneets  1956;
Babuel-Peyrissac  \& Rouvillois  1969; Molodtsov 1994) have introduced into
the equations.  Often kinetic equations (e.g.  radiative transfer equation)
are  deduced  heuristically  from  intuitive  considerations  and  previous
experience.  However,  in the  case  when  polarization  including  induced
effects is considered,  our intuition and experience are not enough and the
phenomenological approach fails.  Therefore, one has to resort to deductive
methods.  Note that in  nearly  all the  works on  Compton  scattering  the
electrons   were   assumed   to   be   unpolarized    and   isotropic   (or
monodirectional).

The first attempt to find out a rule for consideration of induced processes
simultaneously  with  polarization was undertaken by Pomraning  (1974).  In
terms of plane  electromagnetic  waves, he considered a source of polarized
emission  interacting with already  existing  radiation field and deduced a
matrix describing the transformation of the Stokes  parameters.  The Stokes
vector corresponding to the induced process was presented as a product of a
$4\times 4$ matrix $\mN$ with the Stokes vector of the incoming  radiation.
Calculating  the elements of the matrix  $\mN$, he averaged  the product of
the quantities corresponding to the emitted (source photons) and passing by
(field photons) waves over the  distributions of their phases assuming that
no correlation exists between the source and the field.  This result is not
applicable to  scattering  because the  scattered  radiation  does not have
random phases.  In the resulting expression for $\mN$, only the elements in
the first row and the first  column  (i.e.  related to the  intensity)  are
correct in the case of scattering (Stark 1981).

Derivation of the kinetic equation for the induced  scattering of polarized
radiation by  non-relativistic  electrons  was given by Wilson  (1978).  He
used the  Maxwell  equations,  expansion  electromagnetic  field  on set of
harmonic  oscillators, second quantization method and the  non-relativistic
perturbation  theory.  The scattering (a process of the fourth order in the
approximation used) was represented as consequent  absorption and emission.
The products of the  probabilities  of these processes were replaced by the
elements of the scattering matrix deduced according to the rules of quantum
electrodynamics.  While  such  a  deduction  is  not  self-consistent,  the
resulting equation is correct.

Wilson's results were used by Stark (1981) to deduce the  generalization of
the Babuel-Peyrissac---Rouvillois (1969) equation for stimulated scattering
to linearly polarized  radiation (two Stokes  parameters).  Hansen \& Lilje
(1999) have corrected  inaccuracies and misprints in Stark's work.  Using a
version of Wilson's  equation  containing  only  non-linear  terms,  Wilson
(1982) and Coppi et al.  (1993)  showed  that the  induced  scattering  can
enhance  the  polarization   (comparing  to  the  spontaneous   scattering)
producing   also  its   strong   frequency   dependence.  Induced   Compton
backscattering,  for  example,  can amplify an  incident  $\sim 1$ per cent
polarization  up to $\sim 50$ per cent (Sincell \& Coppi 1996).  Accounting
for the  polarization  of radiation  may lead to the results  qualitatively
different from those obtained  neglecting  polarization.  A small change in
the scattering rate (affected by  polarization)  can be strongly  amplified
due to the non-linearity of the process.

The relativistic  kinetic  equation for the Stokes  parameters  taking into
account the induced  scattering  was presented by Nagirner  (1994)  without
deduction.  The work of Wilson was not  accessible  for the  author at that
time.  Shortly  after that, the work of Ioffe made at the  beginning of the
1950-ies was published (Ioffe 1994).  In this work the equation was deduced
(as Ioffe pointed out, with an approximate and a simplified method) for the
so called polarization tensor which differs from the density matrix because
the unphysical scalar and longitudinal photons are not excluded.

In this paper, we derive the relativistic  kinetic  equation for the photon
gas  interacting  with the electron  gas.  The way of deduction is based on
the methods of Bogoliubov  (Bogoliubov \& Gurov 1947) and Baranger  (1958).
The  scheme   closest  to  ours  is  used  in  Silin   (1971),   where  the
non-relativistic  kinetic  equation for interacting  electrons was derived.
In our  paper, at every  step of the  derivation  all the  expressions  and
relations  are  demonstrated  to  be  either  explicitly   relativistically
covariant  or  they  can  be  changed  to  relativistic  forms.  The  final
equations are explicitly  relativistic.  In order to elucidate  these facts
and to  introduce  suitable  for our  purposes  notations,  we are urged to
present a number of known  relations  and schemes.  We thus fully  describe
the method of the derivation of the equations in a self-contained way.

We make assumptions  usual for the kinetic theory.  Only binary  collisions
are accounted for.  We assume the molecular chaos, i.e.  the gas states are
characterized by one-particle  distribution  functions.  The characteristic
temporal and space  scales of a single  interaction  are assumed to be much
smaller than the scales of  significant  changing of radiation and electron
gas states.  We use the principle of weakening of the  correlations,  which
says that the correlations  between  particle states are expressed in terms
of the  same  one-particle  distribution  functions.  There  are  no  other
limitations on the states of  interacting  particles.  Simultaneously  with
the  kinetic  equations,  we  obtain  the  expressions  for the  scattering
amplitudes and (well known) cross-sections.

We first  assume that both  electrons  and photons are  polarized,  then we
average the equations over electron polarizations.  An arbitrary anisotropy
and  nonhomogeneity  of the radiation field and electron  distribution  are
allowed.  We also derive the kinetic  equation  describing the evolution of
the electron gas.

\section{Density Matrix of the Photon-Electron Gas}

\subsection{Operators of Creation and Annihilation of Photons}

\label{sec:crannpho}
According to the method of second quantization (see e.g. Bogoliubov \&
Shirkov 1959; Schweber 1961), the two types of vector-operators
are introduced satisfying the following commutation relations
\be \label{eq:transrel}
[a_\mu(\vk),\ovl{a}_\nu(\vkpr)]=a_\mu(\vk)\ovl{a}_\nu(\vkpr)-
\ovl{a}_\nu(\vkpr)a_\mu(\vk)=-g_{\mu\nu}k\delta(\vk-\vkpr), \quad
\mu,\,\nu=0,1,2,3,
\ee
where $\vk$ is the photon momentum, $k=|\vk|$ and the metrics
$\{g_{\mu\nu}\}=\diag  \{1,-1,-1,-1\}$.  The commutators of all other
components are equal to zero, i.e.
$[\amu(\vk),\anu(\vkpr)]=[\ovl{a}_\mu(\vk),\ovl{a}_\nu(\vkpr)]=0$. The
commutator of the zeroth components of the vector-operator, $a_0(\vk)$ and
$\ovl{a}_0(\vkpr)$, in equation~(\ref{eq:transrel}) differs from others.
Operators  $a$  and  $\ovl{a}$  are interpreted  as  the  operators  of
annihilation  and  creation  of  a  photon,   respectively.  Their  product
$\ovl{a}_{(\lambda)}(\vk)  a_{(\lambda)}(\vk)$   (no   summation)  is  the
operator  of the  number of  photons  of  momentum  $\vk$ and  polarization
$\lambda$.

Generally speaking, one has to quantize the photon field in a finite volume
$V$ (box), for example, of the parallelepiped shape with the sides equal to
$\Lx, \Ly, \Lz,\,V=\Lx\ \Ly\ \Lz $.  Then in every dimension of this volume
there  can  be  a  countable   number  of   standing   waves  of  the  form
$(2\pi\hbar)^{-3/2}\exp(i\vk\vr/\hbar)$,   where  $\vk=2\pi\hbar  (\nx/\Lx,
\ny/\Ly,  \nz/\Lz)$ and $\nx,\ny,\nz$  are integers.  Correspondently,  the
number of creation and annihilation  operators  should be countable and the
commutation  rules should contain  discrete  $\delta$-functions,  i.e.  the
Kronecker  symbols.  However,  in  the  limit   $\Lx,\Ly,\Lz\to\infty$  the
countable  set of  standing  waves  transforms  to the  continuum  and  all
relations for the finite  volume and its limiting  case can be written in a
unified way (Bogoliubov \& Shirkov 1959).

For example, the orthogonality condition for the discrete case with the
integration over the box has the same form as in the continuum case, if the
$\delta$-function and the volume element of the three-dimensional  momentum
space are taken in the discrete form:
\be  \label{eq:boxdel}
\frac{1}{(2\pi\hbar)^3}\int\exp\{i(\vk-\vkpr)\vr/\hbar\}\d^3r=
\delta(\vk-\vkpr),\quad
\delta(\vk-\vkpr)=\frac{V}{(2\pi\hbar)^3}\delta_{\nx\nx'}\delta_{\ny\ny'}
\delta_{\nz\nz'}, \quad \d^3k=\frac{(2\pi\hbar)^3}{V}.
\ee
Therefore, we keep nearly all the quantities  and relations in the form
corresponding to the continuum, while they are valid for the  discrete  case as
well (i.e., for a finite  box).

Let us expand vectors $\ua(\vk)=\{a_0(\vk), \va(\vk)\}= \{a^\mu(\vk)\}$ and
$\ovl{\ua}(\vk)$ along the unit vectors of the four-dimensional basis whose
vectors are orthonormal and their system is full:
\be \label{eq:uaue}
\ua(\vk)=a^{(\lambda)}(\vk)\ue_{(\lambda)},\quad
\ovl{\ua}(\vk)=\ovl{a}^{(\lambda)}(\vk)\ue_{(\lambda)},\quad
\ue_{(\lambda)}\ue_{(\lambda')}=e_{(\lambda)\mu}e^{\mu}_{(\lambda')}=
g_{\lambda\lambda'}, \quad \,e_{(\lambda)\mu}e^{(\lambda)}_\nu=g_{\mu\nu}
\ee
(hereafter, summation over repeated indices is assumed). Raising and lowering
the indices in brackets numbering the unit vectors are made in the same way as
for the  case of  ordinary  indices  numbering  the vector components.  The
coefficients of the expansion, $a^{(\lambda)}(\vk)$ and
$\ovl{a}^{(\lambda)}(\vk)$,  are Lorentz  invariants  (i.e.,  they are
scalars).  They  satisfy  the  commutation   relations  which  follow  from
equations (\ref{eq:transrel}) and (\ref{eq:uaue})
\be \label{eq:transal}
[a^{(\lambda)}(\vk), \ovl{a}^{(\lambda')}(\vkpr)] =-g^{\lambda\lambda'}k
\delta(\vk-\vkpr), \quad \lambda,\lambda'=0, 1, 2, 3.
\ee
In order for the selected basis to become a polarization  basis, their unit
vectors should satisfy the additional  relations  that the two unit vectors
corresponding  to the transverse  polarization  should be orthogonal to the
photon momentum
\be \label{eq:ortkmom}
\uk\,\ue_{(1)}(\vk)=\uk\,\ue_{(2)}(\vk)=0
\ee
where $\uk=\{k,\vk\}$ is the photon four-momentum and the scalar product
$\uk\,\ue=k^\mu e_\mu=ke_0-\vk\cdot\ve$. As a result of such orthogonality, the
photon momentum can be represented as a linear combination of the two unit
vectors of the polarization basis
\be \label{eq:klincomb}
\uk=[\uk\,\ue_{(0)}(\vk)][\ue_{(0)}(\vk)+\ue_{(3)}(\vk)].
\ee
Then the following equality holds
\be \label{eq:ukua}
\uk\,\ua(\vk)=[\uk\, \ue_{(0)}(\vk)] [ a^{(0)}(\vk)-a^{(3)}(\vk)],
\ee
and similarly for the conjugate  operator.  Obviously,  the unit vectors of
the  polarization  basis are not arbitrary but related to the vector of the
photon    momentum    $\vk$.    However,    the    commutation    relations
(\ref{eq:transal}) are still valid.

In classical  electrodynamics,  the  transversality  relation  was given by
equality to zero of the quantity (not an operator) formally coinciding with
(\ref{eq:ukua}).  The equality of the amplitudes of scalar and longitudinal
potentials  (or  equality  of both to  zero)  was the  consequence  of that
relation.  In quantum  electrodynamics,  this  relation  is too stiff.  One
cannot require equality of $a^{(0)}(\vk)$  and  $a^{(3)}(\vk)$,  since they
satisfy  different  commutation  relations  which  easily  can be seen from
relation  (\ref{eq:transal}).  The  solution of that  problem  was given by
Gupta  (1950) and Bleuler  (1950)  (see also  Bogoliubov  \& Shirkov  1959;
Schweber 1961).  We use their method.

Only the operators with indices 1 and 2 are physical and only those should
enter the quantities having physical meaning. These operators satisfy the
commutation relations
\be\label{eq:transas}
[a^{(s)}(\vk),\ovl{a}^{(s')}(\vkpr)] = [ a_{(s)}(\vk),\ovl{a}_{(s')}(\vkpr)]
=\delta_{ss'}k\delta(\vk-\vkpr), \quad s, s'= 1,2.
\ee

\subsection{Photon States}

The   states  of  the   electromagnetic   field  in  the   second-quantized
representation  are  described by the vectors with the  arguments  equal to
given  momenta  and  polarization  projections  of the  definite  number of
photons.  The initial  vector  that is used to obtain all other  vectors is
the state  vector of the  photon  vacuum  $\Psi_0$.  It is  normalized  and
satisfies the condition of absence of photons: $\ovl{\Psi}_0\Psi_0=1,\,
a^{(\lambda)}(\vk)\Psi_0=0$. The second equality is valid for any momentum
$\vk$ and any polarization $\lambda$.

All other photon state vectors can be obtained from the vacuum state vector
by applying the creation operators but the  vectors  containing  information
about the  non-physical  scalar and longitudinal  photons  should be  excluded.
In the  physical  states,  the operators of photon  creation of these  photons
should  appear only in the combination
$\ovl{a}^{(0)}(\vk)+\ovl{a}^{(3)}(\vk)$,  so that there always exist equal
amount of scalar and longitudinal  photons.  With the action of the  operator
${a}^{(0)}(\vk)-{a}^{(3)}(\vk)$  on  to  such  states,  such photons  mutually
cancel out.  The product of the operators of creation and annihilation
in such a combination gives
$\ovl{a}^{(0)}(\vk){a}^{(0)}(\vk)-\ovl{a}^{(3)}(\vk){a}^{(3)} (\vk)$, which
acting  on to the  vector  of the  physical  states  gives  zero.  The full
product of the operators of creation and annihilation is reduced to
$\ovl{\ua}(\vk)\ua(\vk)=-\ovl{a}_{(s)}(\vk)a_{(s)}(\vk)$.
Here the summation is done over $s=1, 2$ (in the several following sections we
use only lower indices $s$). We omit unphysical photons  hereafter  from our
considerations.

The physically allowed states of $N$ photons with fixed momenta and
polarizations can be represented in the form
\be \label{eq:allowstates}
\Psi_{s_1... s_N}(\vk_1,...,\vk_N)=\frac{1}{\sqrt{N!}}
\ovl{a}_{(s_1)}(\vk_1)...\ovl{a}_{(s_N)}(\vk_N) \Psi_0.
\ee
The related arguments $s$ and $\vk$ can be moved simultaneously and rearranged
in any order, since operators commutate.

The orthogonality condition for the vectors of physical states follows from the
commutation relations (\ref{eq:transas}) and takes the form
\be \label{eq:ortoPsi}
\ovl{\Psi}_{s_1'... s_N'}(\vk_1',...,\vk_N')
\Psi_{s_1... s_N}(\vk_1,...,\vk_N)= \frac{1}{N!}
\sum_{\alpha_1,...,\alpha_N}
\prod_{j=1}^N[k_j\delta_{s_j s_{\alpha_j}'}\delta(\vk_j-\vk_{\alpha_j}')],
\ee
where  the sum is taken  over all  permutations  $\alpha_1,...\alpha_N$  of
indices  $1,2,...,N$.  These indices can  alternatively  be assigned to the
arguments  without  primes, which is equivalent  to the change from the sum
over  rows to the sum over  columns.  

The results of the action of the creation and annihilation operators on to the
state vector are given by
\beq \label{eq:aPsi}
a_{(s)}(\vk)\Psi_{s_1... s_N}(\vk_1,..., \vk_N)&=&
\frac{1}{\sqrt{N}} \sum_{r=1}^N \ [ k\delta_{s s_r}\delta(\vk\!-\!\vk_r) ]
\ \Psi_{s_1...s_{r-1}s_{r+1}...s_N}
(\vk_1,...,\vk_{r-1},\vk_{r+1},...,\vk_N),   \\
\ovl{a}_{(s)}(\vk)\Psi_{s_1...s_N}(\vk_1,...,\vk_N)&=&
\sqrt{N+1}\ \Psi_{ss_1...s_N}(\vk,\vk_1,...,\vk_N).
\eeq
The product of these operators acts on to the vector as follows
\be \label{eq:aaPsi}
\ovl{a}_{(s)}(\vk)a_{(s')}(\vkpr)\Psi_{s_1...s_N}(\vk_1,...,\vk_N)\!=\!
\sum_{r=1}^Nk'\delta_{s's_r}\delta(\vkpr\!-\!\vk_r)
\Psi_{s_1...s_{r-1}ss_{r+1}... s_N}
(\vk_1,...,\vk_{r-1},\vk,\vk_{r+1},...,\vk_N).
\ee

Arbitrary physical state with $N$ transverse photons can be described by 
the vector
\be \label{eq:Psiphys}
\Psi_N=\int\frac{\d^3k_1}{k_1}...\frac{\d^3k_N}{k_N}
c_{s_1...s_N}(\vk_1,...,\vk_N)\Psi_{s_1...s_N}(\vk_1,...,\vk_N) ,
\ee
where  the  sum  is  taken  over  the  repeated   indices   $s_j=1,2$.  The
coefficients  of the expansion can be found as usual Fourier  coefficients.
We  assume  that  vectors  of the  physical  state  (\ref{eq:Psiphys})  are
dimensionless.  They are normalized as $\ovl{\Psi}_N\Psi_N=1$.  In order to
prove that, the orthogonality  condition  (\ref{eq:ortoPsi}) should be used
and the  integral  over  momenta  $\vkpr_i$  must  be  taken  applying  the
permutation property for indicies and momenta.

\subsection{Quantization of the Electron-Positron Field}

In the  second-quantization  of the electron and positron fields (as in the
case of the  electromagnetic  field),  the operators of annihilation and
creation  satisfy the relations of the form
\be \label{eq:tranrel}
\{b_\tau(\vp),b_{\tau'}^\dagger(\vppr)\}=b_\tau(\vp)
b_{\tau'}^\dagger(\vppr)+b^\dagger_{\tau'}(\vppr)b_\tau(\vp)=
\delta_{\tau\tau'}p_0\delta(\vp-\vppr),\,\,\mbox{for electrons},
\ee
and  the  same  for  the  positron   operators   $d_\sigma(\vp)$.  Relation
(\ref{eq:tranrel}) (and a similar one for positrons) contains anticommutators
(not commutators),  which is a consequence of the fact that these particles
are fermions and satisfy the Pauli  exclusion  principle.  All other binary
combinations of the operators anticommutate.

The electron-positron vacuum is defined with normalized vector $\Phi_0$ as
a state where there are no particles, i.e.
\be \label{eq:Phi0}
b_\tau(\vp)\Phi_0=0, \quad d_\sigma(\vp)\Phi_0=0,\quad \Phi^\dagger_0\Phi_0=1.
\ee
Then the state with $\Ne$ electrons can be defined via vector
\be \label{eq:Phitspq}
\Phi_{\tau_1...\tau_{\Ne}}(\vp_1,...,\vp_{\Ne})=\frac{1}{\sqrt{\Ne!}}
b^\dagger_{\tau_1}(\vp_1)...b^\dagger_{\tau_{\Ne}}(\vp_{\Ne})\Phi_0.
\ee
The  order  of  the  indices  and  corresponding  momenta  can  be  changed
arbitrarily,  but since the operators now  anti-commutate,  the sign of the
vector  changes  depending  whether  a  permutation  is odd  or  even.  The
positron  states can be defined in a similar  manner.  Since the scattering
of photons by positrons  does not  substantially  differ from the  electron
case, we can assume that it is only  electrons  that interact  with photons
and we do not introduce positron states.

The operators of creation and annihilation act on to the state vectors in 
the following way
\be \label{eq:bPhi}
 b_\tau(\vp)\Phi_{\tau_1...\tau_{\Ne}}\!(\vp_1,...,\vp_{\Ne})\!
=\!\!\frac{1}{\sqrt{\Ne}}\sum_{j=1}^{\Ne}(-1)^{j-1}\delta_{\tau\tau_j}
p_0\delta(\vp\!-\!\vp_j)\Phi_{\tau_1...\tau_{j-1}\tau_{j+1}...\tau_{\Ne}}
(\vp_1,...,\vp_{j-1}\vp_{j+1},...,\vp_{\Ne}),
\ee
\be\label{eq:bdPhi}
b^\dagger_\tau(\vp)\Phi_{\tau_1...\tau_{\Ne}}
(\vp_1,...,\vp_{\Ne})=\sqrt{{\Ne}+1} \ \Phi_{\tau\tau_1...\tau_{\Ne}}
(\vp,\vp_1,...,\vp_{\Ne}).
\ee
The products of operators give the following
\be \label{eq:bdbPhi}
b^{\dagger}_{\tau}(\vp)b_{\tau'}(\vppr)
\Phi_{\tau_1...\tau_{\Ne}} (\vp_1,...,\vp_{\Ne})=
\sum_{j=1}^{\Ne}\delta_{\tau'\tau_j}p'_0\delta(\vppr-\vp_j)
\Phi_{\tau_1...\tau_{j-1}\tau\tau_{j+1}...\tau_{\Ne}}
(\vp_1,...,\vp_{j-1},\vp,\vp_{j+1},...,\vp_{\Ne}).
\ee

Normalization  of the state  vectors is defined in a manner  similar to  
photons.  One has to change  the order of the  creation  and
annihilation  operators  many  times in order for the  former  to be in the
beginning and for the later to be in the end.  The only  difference is that
a minus sign appears for odd permutation.  The result can be written in the
determinant form.  For electrons,
\be \label{eq:PhiPhiel}
\Phi^\dagger_{\tau'_1...\tau'_{\Ne}}(\vppr_1,...,\vppr_{\Ne})
\Phi_{\tau_1...\tau_{\Ne}}(\vp_1,...,\vp_{\Ne})=\frac{1}{{\Ne}!}\det\left(
\delta_{\tau_i\tau'_j}p_{i0}\delta(\vp_i-\vppr_j)\right).
\ee

The vector of an arbitrary state with ${\Ne}$ electrons can be
represented via an expansion over the elementary state vectors
\be \label{eq:Phielpos}
\Phi_{\Ne}=\int\frac{\d^3p_1}{p_{10}}...\frac{\d^3p_\Ne}{p_{{\Ne} 0}}
c_{\tau_1...\tau_{\Ne}}(\vp_1,...,\vp_{\Ne})\Phi_{\tau_1...\tau_{\Ne}}
(\vp_1,...,\vp_{\Ne}).
\ee
The interchange of the arguments of the coefficients $c$ changes their sign.
The normalization of the state vectors, $\Phi^\dagger_{\Ne}\Phi_{\Ne}=1$, 
can  be easily obtained from equations (\ref{eq:PhiPhiel}) and 
(\ref{eq:Phielpos}).

\subsection{Density Matrices for Photons and Electrons}

The density matrix of $N$-photon state ($N$ is fixed) can be defined as an
averaged dyad product of the state vector with its conjugate
\beq \label{eq:rho}
 \rho _N &=&\langle\Psi_N\ovl{\Psi}_N\rangle \nonumber \\
&=&\int\frac{\d^3k'_1}{k'_1}...\frac{\d^3k'_N}{k'_N}
\frac{\d^3k_1}{k_1}...\frac{\d^3k_N}{k_N}\langle c^*_{s'_1...s'_N}
(\vkpr_1,...,\vkpr_N)c_{s_1...s_N}(\vk_1,...,\vk_N)\rangle
\Psi_{s_1...s_N}(\vk_1,...,\vk_N)
\ovl{\Psi}_{s'_1...s'_N}(\vkpr_1,...,\vkpr_N).
\eeq
This product is an operator, since the elementary state vectors contain the
creation and annihilation operators.

Let us introduce the notation for the kernel of the density matrix operator
\be \label{eq:rhokern}
\rho^{s'_1...s'_N}_{s_1...s_N}\biggl({\vkpr_1\atop\vk_1}{...\atop...}
{\vkpr_N\atop\vk_N}\biggr)=N!\ \langle c^*_{s'_1...s'_N}(\vkpr_1,...,\vkpr_N)
c_{s_1...s_N}(\vk_1,...,\vk_N)\rangle.
\ee
This kernel, being the $N$-particle photon distribution function in the
momentum representation, cannot be represented as a product after averaging
of the expansion coefficients $c$. The density matrix is then written in the
form
\be \label{eq:rhorhokern}
\rho_N=\frac{1}{N!}\int\frac{\d^3k'_1}{k'_1}...\frac{\d^3k'_N}{k'_N}
\frac{\d^3k_1}{k_1}...\frac{\d^3k_N}{k_N}\rho^{s'_1...s'_N}_{s_1...s_N}
\biggl({\vkpr_1\atop\vk_1}{...\atop...}{\vkpr_N\atop\vk_N}\biggr)
\Psi_{s_1...s_N}(\vk_1,...,\vk_N)\ovl{\Psi}_{s'_1...s'_N}(\vkpr_1,...,\vkpr_N),
\ee
and the kernel can be represented through the operator
\be \label{eq:rhokernrho}
\rho^{s'_1...s'_N}_{s_1...s_N}\biggl({\vkpr_1\atop\vk_1}{...\atop...}
{\vkpr_N\atop\vk_N}\biggr)=N!\ \ovl{\Psi}_{s_1...s_N}(\vk_1,...,\vk_N)
\rho_N\Psi_{s'_1...s'_N}(\vkpr_1,...,\vkpr_N).
\ee
Only diagonal elements (i.e., elements with the same primed and non-primed
indices and arguments) are real and non-negative.

It is obvious from definition  (\ref{eq:rho}), that the density matrix is a
self-conjugate   operator,  and  it  follows  from  (\ref{eq:rhokern})  and
(\ref{eq:rhorhokern}) that its kernel is a self-conjugate matrix, i.e.  its
Hermitian conjugation (complex conjugation and the replacement of the lower
and the upper arguments with each other) is equal to the same kernel.
The trace of the density matrix $\Sp\rho_N=\langle\ovl{\Psi}_N\Psi_N\rangle=1$
by virtue of the normalization of the state vector.

The  density  matrix  carries  a lot of  information,  much  more  than the
distribution  function  or  even  the  polarization  matrix  which  can  be
expressed  through  the  density  matrix.  However,  in order to derive the
kinetic equation, the distribution  functions are not enough and one has to
introduce a set of functions  containing  the groups of variables  with the
dimension less than $N$, but larger than 1.  We call them truncated density
matrices.

Together with the $N$-particle density matrix we introduce matrices whose
kernels can be expressed through the integrals of the kernel of the original
matrix. A kernel of order $l$
\be \label{eq:rhol}
\rho^{s_1'...s_l'}_{s_1...s_l}\biggl({\vk_1'\atop\vk_1}{...\atop...}
{\vk_l'\atop\vk_l}\biggr)=\frac{1}{(N-l)!}\int\frac{\d^3k_{l+1}}{k_{l+1}}...
\frac{\d^3k_N}{k_N}\rho^{s_1'...s_l's_{l+1}...s_N}_{s_1...s_ls_{l+1}...s_N}
\biggl({\vk_1'\atop\vk_1}{...\atop...}{\vk_l'\atop\vk_l}
{\vk_{l+1}\atop\vk_{l+1}}{...\atop...}{\vk_N\atop\vk_N}\biggr)
\ee
corresponds to the truncated $l$-particle matrix. Integrals (\ref{eq:rhol})
are normalized according to normalization of the matrix $\rho_N$, so that
\be \label{eq:rholnorm}
\int\frac{\d^3k_1}{k_1}...\frac{\d^3k_l}{k_l}\rho^{s_1...s_l}_{s_1...s_l}
\biggl({\vk_1\atop\vk_1}{...\atop...}{\vk_l\atop\vk_l}\biggr)=
\frac{N!}{(N-l)!}=N(N-1)...(N-l+1).
\ee
Specifically, one-particle photon density matrix
\be \label{eq:rho1}
\rho^{s'}_s\biggl({\vkpr\atop\vk}\biggr)=\frac{1}{(N-1)!}\int
\frac{\d^3k_2}{k_2}...\frac{\d^3k_N}{k_N}\rho^{s's_2...s_N}_{ss_2...s_N}
\biggl({\vkpr\atop\vk}{\vk_2\atop\vk_2}{...\atop...}{\vk_N\atop\vk_N}\biggr)
\ee
is normalized to the number of particles
\be \label{eq:rho1N}
\int\frac{\d^3k}{k}\rho^s_s\biggl({\vk\atop\vk}\biggr)=N.
\ee

The action of the product of the annihilation and creation operators on to
the density matrix is reduced to
\beq \label{eq:aarho}
 \ovl{a}_{(s)}(\vk)a_{(s')}(\vkpr)\rho &=&
\int\frac{\d^3k_1'}{k_1'}...\frac{\d^3k_N'}{k_N'}\frac{\d^3k_1}{k_1}...
\frac{\d^3k_N}{k_N}\sum_{j=1}^Nk'\delta_{s's_j}\delta(\vkpr-\vk_j)
  \nonumber \\
&\times & \Psi_{s_1...s_{j-1}ss_{j+1}...s_N}(\vk_1,...,\vk_{j-1},
\vk,\vk_{j+1},...,\vk_N)\ovl{\Psi}_{s_1'...s_N'}(\vkpr_1,...,\vkpr_N)
\rho^{s_1'...s_N'}_{s_1...s_N}\biggl({\vkpr_1\atop\vk_1}{...\atop...}
{\vkpr_N\atop\vk_N}\biggr).
\eeq
Therefore, we get
\beq \label{eq:Psiaarho}
 \ovl{\Psi}_{s_1...s_N}(\vk_1,...,\vk_N)\ovl{a}_{(s)}
(\vk)a_{(s')}(\vkpr)&\rho&
\Psi_{s'_1...s'_N}(\vkpr_1,...,\vkpr_N)  \nonumber \\
&=&\frac{1}{N!}\sum_{j=1}^Nk\delta_{ss_j}
\delta(\vk-\vk_j)\rho^{s'_1...s'_{j-1}s'_js'_{j+1}...s'_N}
_{s_1...s_{j-1}s's_{j+1}...s_N}\biggl({\vkpr_1\atop\vk_1}{...\atop...}
{\vkpr_{j-1}\atop\vk_{j-1}}{\vkpr_j\atop\vkpr}{\vkpr_{j+1}\atop\vk_{j+1}}
{...\atop...}{\vkpr_N\atop\vk_N}\biggr).
\eeq

The density matrices for particles  (electrons) are introduced in a similar
manner,  therefore we immediately  write down the expression  for the joint
density matrix for electrons and photons.  It can be expressed  through the
kernel in terms of the following integral
\beq \label{eq:kernrhorho}
 \rho&=&\frac{1}{N!{\Ne}!}\int\frac{\d^3k_1}{k_1}
\frac{\d^3k_1'}{k_1'}...\frac{\d^3k_N}{k_N}\frac{\d^3k_N'}{k_N'}\frac{\d^3p_1}
{p_{10}}\frac{\d^3p_1'}{p_{10}'}...\frac{\d^3p_{\Ne}}{p_{{\Ne}0}}\frac
{\d^3p'_{\Ne}}{p_{{\Ne}0}'}
\rho^{s_1'...s_N',\tau_1'...\tau_{\Ne}'}_{s_1...s_N,
\tau_1...\tau_{\Ne}}\biggl({\vk_1'\atop\vk_1}{...\atop...}
{\vk_N'\atop\vk_N}\left|{\vppr_1\atop\vp_1}{...\atop...}
{\vppr_{\Ne}\atop\vp_{\Ne}}\right.\biggr)   \nonumber \\
& \times & \Psi_{s_1...s_N}(\vk_1,...,\vk_N)
\ovl{\Psi}_{s_1'...s_N'}(\vk_1',...,\vk_N')
\Phi_{\tau_1...\tau_{\Ne}}(\vp_1,...,\vp_{\Ne})
\Phi^\dagger_{\tau_1'...\tau_{\Ne}'}(\vppr_1,...,\vppr_{\Ne}).
\eeq

Changing the order of the photon arguments does not change  anything, while
changing  the order of electron  arguments  in the state  vectors or in the
kernel of the density  matrix  changes  the sign.  However,  if one changes
simultaneously  arguments  in the state  vectors as well as in the  kernel,
nothing  changes  since  this  procedure  corresponds  to just a change  of
notations.  The sign does not  change if one  interchange  lower  and upper
arguments, since the sign here changes even number of times.  The truncated
density  matrix of electrons and their kernels can be introduced  following
the same  procedure as in the case of photons.  All the  relations  for the
photon matrices are valid for the electrons too.

\subsection{One-particle Distribution Functions}

The  kinetic  equation  for  photons  which  we wish to  deduce  should  be
formulated for the one-particle  polarization  matrix  depending on spatial
coordinates,  time,  and  photon  momentum.  Therefore,  one has to  make a
transformation  to such a matrix.  We briefly  describe the scheme for this
transformation on the example of spinless particles.

Up to now we used the momentum  representation  where the density  matrices
depend on a double set of momenta  and  polarization  indices.  The  spinless
states could be described in the coordinate representation.  Instead of the
matrix with two momenta  arguments, we would have a matrix with two sets of
space  coordinates.  The transition from one  representation to another can
be performed by Fourier  transform.  For the one-particle  functions such a
transition is defined by the formula
\be \label{eq:rrrpp}
\rho_{\rm s}(\vr,\vrpr)=\frac{1}{(2\pi\hbar)^6}\int\frac{\d^3p\,\,\d^3p'}
{\sqrt{p_0p_0'}}e^{-i(\vps\vrs-\vps'\vrs')/\hbar}
\rho\biggl({\vppr\atop\vp}\biggr),
\ee
where $\ur=\{ct,\vr\}$. The inverse transform is
\be \label{eq:rpprr}
\rho\biggl({\vppr\atop\vp}\biggr)=
\sqrt{p_0p_0'} \int\d^3r\d^3r' e^{i(\vps\vrs-\vps'\vrs')/\hbar}
\rho_{\rm s}(\vr,\vrpr).
\ee
The two factors under the root are introduced since it is convenient when
transforming to our notations. In the kinetic theory, the transition from
the matrix with two space arguments to the usual distribution function is
done via the Wigner function (Wigner 1932; see also Silin 1971), which is
defined as follows
\be \label{eq:vig1}
\rho(\vp,\vr)=\int\d^3v e^{i\vps\vvs/\hbar}\rho_{\rm s}(\vr+\vv/2,\vr-\vv/2).
\ee
The Wigner function can be also expressed through the matrix with two momentum
arguments
\be \label{eq:Wigrpp}
\rho(\vp,\vr)=\frac{1}{(2\pi\hbar)^6}\int\frac{\d^3p_1\d^3p_1'}
{\sqrt{p_{01}p_{01}'}}e^{-i(\vps_1-\vps'_1)\vrs/\hbar}\delta
\left(\vp-\frac{\vp_1+\vppr_1}{2}\right)\rho\biggl({\vppr_1\atop\vp_1}\biggr).
\ee
The inverse transform reads
\be \label{eq:vig2}
\rho\biggl({\vppr_1\atop\vp_1}\biggr)=
\frac{\sqrt{p_{01}p_{01}'}}{(2\pi\hbar)^3}
\int\d^3p\d^3re^{i(\vps_1-\vps'_1)\vrs/\hbar}
\delta\left(\vp-\frac{\vp_1+\vppr_1}{2}\right)\rho(\vp,\vr).
\ee
According to the assumption of the small scale of the  interaction  between
photons and  electrons  as compared  with the  macroscopic  scale where the
distribution function in  equation~(\ref{eq:vig2})  changes  significantly,
one can  assume  that the  Wigner  function  does not  depend  on the space
coordinates.  Then both integrals in equation~(\ref{eq:vig2})  can be taken
and the matrix becomes diagonal in momenta
\be \label{eq:vigdia}
\rho \biggl({\vppr_1\atop\vp_1}\biggr)=p_{01}\delta(\vppr_1-\vp_1)
\rho(\vp_1).
\ee

Here we  deduced  these  expression  in  three-dimensional  coordinate  and
momentum  space,  i.e.  non-relativistically.  In order  to  introduce  the
relativistic  generalization  of the Wigner  function (see, e.g., de Groot,
van Leeuwen, \& van Weert  1980), one has to define it for the  generalized
momentum $\up$ which does not satisfy the relation  $\up^2=p_0^2-\vp^{\,2}=
m^2c^2$.  Then one can prove that  outside of the  surface  defined by this
relation in the  four-dimensional  momentum  space, the  function  strongly
oscillates    (Zitterbewegung)   and,   on   average,   vanishes.   Further
consideration leads to equation~(\ref{eq:vigdia}).

In the case of photons, situation becomes even more complicated,  since the
physical  states with the  transverse  polarizations  are  connected to the
momentum  and, for the photon  matrices  with the fixed  spin  projections,
transformations of the type (\ref{eq:rrrpp}) and (\ref{eq:Wigrpp})  are not
possible.  In that  case, one has to  return  to the  general  polarization
states which include  non-physical  photons.  The  elimination of the later
should   be  done   after   the   transition   to   equation   similar   to
(\ref{eq:vigdia}).  Here we also meet with the  difficulties to satisfy the
relation $\uk^2=0$.

Similar  difficulty  appears also in the case of electrons if one takes the
solutions with a given spirality  (i.e., a projection of the spin on to the
momentum) as the basic solutions of the Dirac equations.  In that case, the
transformations of the type (\ref{eq:rrrpp}) and (\ref{eq:Wigrpp}) also are
not possible.  However, one can get solutions with the  projections  of the
spin on to an  arbitrary  direction  instead of the  momentum.  This can be
achieved by the method  proposed by Foldy \&  Wouthuysen  (1950)  (see also
Schweber  1961).  The  mentioned  transformations  are then  possible.  The
transition  to the equation of type  (\ref{eq:vigdia})  is done in the same
way as for the spinless  particles  and the spin indices are simply  added.
Without making these derivation, we take that
\be \label{eq:rhottpp}
\rho^{\tau'}_\tau\biggl({\vppr_1\atop\vp_1}\biggr)=p_{01}\delta(\vppr_1-\vp_1)
\rho^{\tau'}_\tau(\vp_1),
\ee
where the indices characterize the spirality. For photons we make a similar
approximation
\be \label{eq:rhoss}
\rhotp{s'}{s}{\vkpr}{\vk}=k\delta(\vkpr-\vk)\rho^{s'}_{s}(\vk),
\ee
The matrices $\rho^{\tau'}_{\tau}(\vp)$ and $\rho^{s'}_{s}(\vk)$ depend, of
course, on time and space coordinates, but this macroscopic
dependence is unimportant for the description of a single scattering act.

In the  normalization  condition  (\ref{eq:rho1N})  the  momenta  $\vk$ and
$\vkpr$  are  equal  to  each  other.  Formally,  for  equal   momenta  the
$\delta$-function in equation  (\ref{eq:rhoss}) becomes infinite.  However,
since we operate with the  quantities  in the finite  volume  (basic  box),
according to (\ref{eq:boxdel}) one has to take
\be \label{eq:deltadisc}
\delta(\vk-\vk)=\frac{V}{(2\pi\hbar)^3}, \quad
\rhotp{s'}{s}{\vk}{\vk}=k\frac{V}{(2\pi\hbar)^3} \rho^{s'}_{s}(\vk).
\ee
Then the normalization condition (\ref{eq:rho1N}) reads
\be \label{eq:normrhok}
\int\frac{\d^3k}{k}\rhotp{s}{s}{\vk}{\vk}=N=
\frac{V}{(2\pi\hbar)^3}\int\d^3k\,\rho^{s}_{s}(\vk),
\quad \mbox{so that} \quad
\frac{1}{(2\pi\hbar)^3}\int\d^3k\,\rho^{s}_{s}(\vk)=\frac{N}{V}.
\ee

When  simultaneously  $V\to\infty$  and  $N\to\infty$  while their ratio is
constant,  relation  (\ref{eq:normrhok})  transforms to the usual condition
normalizing       the       polarization        matrix.       The       sum
$\rho^{s}_{s}(\vk)=\rho^{1}_{1}(\vk)+\rho^{2}_{2}(\vk)$  is  equal  to  the
double  mean  occupation  number of the photon  states  $2\rho(\vk)$.  This
agrees with the fact that unpolarized radiation is described by the matrix
\be \label{eq:matrunpol}
\rho^{s'}_{s}(\vk)=\delta_{ss'}\rho(\vk),
\ee
so that $\rho^{s}_{s}(\vk)=2\rho(\vk)$.

The normalization condition for the electron polarization matrix can be
obtained from the equations analogous to (\ref{eq:rho1N}) and
(\ref{eq:boxdel}). The later takes the form $\delta(\vp-\vp)=V/(2\pi\hbar)^3$,
so that
\be \label{eq:normrhop}
\frac{1}{(2\pi\hbar)^3}\int\d^3p\,\rho^{\tau}_{\tau}(\vp)=\frac{\Ne}{V}.
\ee
The unpolarized electrons are characterized by the diagonal matrix
\be \label{eq:unpolele}
\rho^{\tau'}_{\tau}(\vp)=\frac{(2\pi\hbar)^3}{2}\delta_{\tau}^{\tau'}f_\e(\vp),
\ee
where the distribution function in the comoving frame is normalized to the
electron number density
\be \label{eq:normfep}
\int\d^3p f_\e(\vp)=\frac{\Ne}{V}.
\ee
We use the same notation, $\rho$, for both photon and electron matrices, but
they differ by their arguments and indices.

In spite of the fact that equations (\ref{eq:rhottpp}) and (\ref{eq:rhoss})
are deduced in a non-relativistic way they are relativistically  covariant.
Normalization  conditions  (\ref{eq:normrhok}) and  (\ref{eq:normrhop}) are
written in the reference  frames where the photon and electron gases are at
rest.  The  corresponding   relativistically   covariant  equations  in  an
arbitrary frames are
\be\label{eq:relnorms}
\frac{1}{(2\pi\hbar)^3}\int\frac{\d^3k}{k}\uk\rho^{s}_{s}(\vk)=
\frac{N}{V}\langle\uk\rangle,\quad
\frac{1}{(2\pi\hbar)^3}\int\frac{\d^3p}{p_0}\up
\rho^{\tau}_{\tau}(\vp)=\frac{{\Ne}}{V}\langle\up\rangle ,
\ee
where $\langle\uk\rangle$ and $\langle\up\rangle$ are the average photon and
electron momenta in a given frame.

Let us note that the interaction time $T_0$ is also related to the
$\delta$-function. This relation is the relativistic counterpart of
(\ref{eq:boxdel}), i.e. its time-like form
\be \label{eq:deltak}
\delta(k-k)=\frac{cT_0}{2\pi\hbar}.
\ee

\section{Interaction between Photons and Electrons}

\subsection{Equation for the Density Matrix}

Density matrix (\ref{eq:kernrhorho}) satisfies  the quantum Liouville 
equation (Landau \& Lifshitz 1977; Silin 1971; de Groot et al. 1980)
\be \label{eq:eqLiuv}
i\hbar\Dr{\rho(t)}{t}=H(t)\rho(t)-\rho(t)H(t)
\ee
where  $H(t)$  is a  Hamiltonian.  This  equation  can  be  transformed  to
explicitly  covariant  form  if we  proceed  from  the  Tomonaga--Schwinger
equation  (Bogoliubov  \&  Shirkov  1959).  In  perturbation   theory,  the
solution of this  equation  can be written as a series.  It is shown in the
text  books  on  quantum   electrodynamics  that  a  part  of  this  series
corresponding  to  the  scattering  of a  photon  by  an  electron  can  be
represented in the following  form (index CE stands for Compton  scattering
by an Electron)
\be \label{eq:UCEdef}
U^{\rm CE}_{\rm V}(t,t_0)=\intl_{t_0}^t\d t'\intl_{t_0}^t\d t''
\intl_V\d^3r'\intl_V\d^3r''\cS^{\rm CE}(\ur',\ur'')=
\frac{1}{c^2}\intl_\cV\d^4r'\intl_\cV\d^4r''\cS^{\rm CE}(\ur',\ur''),
\ee
where $\cV$ is part of Minkowsky space $ct_0\!\leq x_0\!\leq ct$, $-\Lx/2\!\leq
x\!\leq\Lz/2$, $-\Ly/2\!\leq y\!\leq\Ly/2$,$-\Lz/2\!\leq z\!\leq\Lz/2$, and
\be \label{eq:cSCErr}
\cS^{\rm CE}(\ur',\ur'')=i\frac{e^2c}{(2\pi\hbar)^9}\int\frac{\d^3p}{p_0}
\frac{\d^3p'}{p'_0}\frac{\d^3k}{k}\frac{\d^3k'}{k'}b^\dagger_{\tau'}(\vppr)
b_\tau(\vp)\ovl{a}_{(s')}(\vkpr)a_{(s)}(\vk)\cNatr46{s\ \tau\;}{s'\tau'}{\vk\;}
{\vkpr}{\vp\;}{\vppr}{\ur''}{\ur'\;}.
\ee
In the last expression
\be \label{eq:Nmatrdef}
\cNatr46{s\ \tau\;}{s'\tau'}{\vk\;}{\vkpr}{\vp\;}{\vppr}{\ur''}{\ur'\;}=
\frac{1}{2}\left[
\cMatr46{s\;\tau\;}{s'\tau'}{\vk\;}{\vkpr}{\vp\;}{\vppr}{\ur''}{\ur'\;}+
\cMatr46{s\;\tau\;}{s'\tau'}{\vk\;}{\vkpr}{\vp\;}{\vppr}{\ur'\;}{\ur''}\right]
\ee
and
\beq \label{eq:Mmatrdef}
\cMatr46{s\;\tau\;}{s'\tau'}{\vk\;}{\vkpr}{\vp\;}{\vppr}{\ur''}{\ur'\;}
&=& mc\,\int\d^4p_{\rm v}\ovl{u}^{\tau'}(\vppr)\left[\hat{e}_{(s')}(\vkpr)
\frac{mc+\hat{p}_{\rm v}}{m^2c^2-\up_{\rm v}^2-i0}\hat{e}_{(s)}(\vk)
e^{i(\uk'\ur'-\uk\,\ur'')/\hbar}  \right.   \nonumber \\
&+& \left. \hat{e}_{(s)}(\vk)\frac{mc+\hat{p}_{\rm v}}
{m^2c^2-\up_{\rm v}^2-i0}\hat{e}_{(s')}(\vkpr)e^{i(\uk'\ur''-\uk\,\ur')/\hbar}
\right]e^{i(\up'\ur'-\up\ur'')/\hbar}e^{-i\up_{\rm v}(\ur'-\ur'')/\hbar}
u^{\tau}(\vp).
\eeq
Here  $u^\tau(\vp)$,  the  elementary   solutions  of  the  Dirac  equation
describing   electron,   are  the  columns   containing   four   functions,
$\ovl{u}^\tau(\vp)= [u^\tau(\vp)]^\dagger\gamma^0$ and $\hat{p}=\up\,\ugam=
p_0\gamma^0-\vp\cdot\vgamma=  p_\mu\gamma^\mu$,  where $\gamma^\mu$ are the
Dirac matrices (e.g.  Schweber 1961; Berestetskii,  Lifshitz, \& Pitaevskii
1982).  The scalar product  $\uk\,\ur=kct-\vk\cdot\vr$.  The imaginary term
$i0$  determines  the  definite  rule  of  circuit  of  singularity  in the
denominator.  It   is   obvious   that   matrices   (\ref{eq:cSCErr})   and
(\ref{eq:Nmatrdef}) are symmetric relative to $\ur'$ and $\ur''$.

If  $V\to\infty$,  $t_0\to-\infty$,   $t\to\infty$,  the  electron  Compton
operator  $U^{\rm  CE}_{\infty}  (-\infty,\infty)=S^{\rm  CE}$ contains two
integrals   over   four-dimensional   space   ($\ur'$  and   $\ur''$).  Two
four-dimensional   $\delta$-functions   appear  after   calculating   these
integrals.  One  of  them  disappears   when  we  take  the  integral  over
four-dimensional  momentum of virtual  electron  $\up_{\rm v}$, whereas the
second one reflects the  conservation  laws.  Thus integrals over the whole
time-space ($x_0=ct$)
\be \label{eq:intNintM}
\int\d^4r'\!\int\!\d^4r''\cNatr46{s\ \tau\;}{s'\tau'}{\vk\;}{\vkpr}{\vp\;}
{\vppr}
{\ur''}{\ur'\;}\!=\!\int\!\d^4r'\!\int\!\d^4r''\cMatr46{s\;\tau\;}{s'\tau'}
{\vk\;}
{\vkpr}{\vp\;}{\vppr}{\ur''}{\ur'\;}\!=\!(2\pi\hbar)^8\delta(\up'\!+\!\uk'\!-
\!\up\!-\!\uk)\matr44{s\ \tau\;}{s'\tau'}{\vk\;}{\vkpr}{\vp\;}{\vppr},
\ee
where the amplitude of the process
\be  \label{eq:amplMdef}
\matr44{s\ \tau\;}{s'\tau'}{\vk\;}{\vkpr}{\vp\;}{\vppr}=mc
\ovl{u}^{\tau'}(\vppr)\left[\he_{(s')}(\vkpr)\frac{mc+\hp+\hk}
{m^2c^2-(\up+\uk)^2}\he_{(s)}(\vk)+\he_{(s)}(\vk)\frac{mc+\hp-\hk'}
{m^2c^2-(\up-\uk')^2}\he_{(s')}(\vkpr)\right]u^{\tau}(\vp).
\ee
The amplitude has an obvious property
\be \label{eq:Mst}
\left[\matr44{s\ \tau\;}{s'\tau'}{\vk\;}{\vkpr}{\vp\;}{\vppr}\right]^*
=\matr44{s'\tau'}{s\;\tau\;}{\vkpr}{\vk\;}{\vppr}{\vp\;}.
\ee
Then for matrix $S^{\rm CE}$ the following expression is obtained (Bogoliubov
\& Shirkov 1959)
\be \label{eq:S2CE}
S^{\rm CE}=\frac{e^2}{\hbar c}\frac{i}{2\pi}\int\frac{\d^3k}{k}
\frac{\d^3k'}{k'}\frac{\d^3p}{p_0}\frac{\d^3p'}{p_0'}\delta(\up'+\uk'-\up-\uk)
b^\dagger_{\tau'}(\vp\,')b_\tau(\vp)\ovl{a}_{(s')}(\vk')a_{(s)}(\vk)
 \matr44{s\ \tau\;}{s'\tau'}{\vk\;}{\vkpr}{\vp\;}{\vppr} .
\ee
It is easy to show that the operators $U^{\rm CE}_{\rm V}(t,t_0)$ and
$S^{\rm CE}$ being conjugated only change the sign.

If we omit all terms which are higher than the second order and those which
do not  conserve  the number and/or the quality of particles then the equation
for the photon-electron density matrix describing Compton scattering by
electrons takes the form
\be \label{eq:rhoU2}
\intl_{t_0}^t\Dr{\rho(t')}{t'}\d t'=\rho(t)-\rho(t_0)=
U_{\rm V}^{\rm CE}(t,t_0)\rho(t_0)+\rho(t_0)
{U_{\rm V}^{\rm CE}}^\dagger(t,t_0).
\ee

The derivative in equation (\ref{eq:rhoU2}) describes all variations of the
density matrix which have very different scales (hierarchy of scales).  The
duration and the characteristic scale of the interaction are much less than
the scales of macroscopic  changes of the matrix and this fact is the basis
of the Bogoliubov method of deducing the kinetic equations.  Thus we assume
that the density matrix $\rho$ changes  negligibly  during the interaction.
Following  Baranger (1958), we take the value of the derivative at time $t$
out of the integral in the lhs of equation  (\ref{eq:rhoU2}), i.e.  the lhs
is now  $T_0\d  \rho/\d  t$.  We  write  the  full  derivative  because  it
describes now macroscopic changes.  Thus we get the initial basic equation
\be \label{eq:rhoUCE}
T_0\DR{\rho(t)}{t}=\rho(t)-\rho(t_0)=U^{\rm CE}_{\rm V}(t,t_0)\rho(t_0)-
\rho(t_0)U^{\rm CE}_{\rm V}(t,t_0).
\ee
  
\subsection{Equation for the Kernel of the Density Matrix}

Let us write equation  (\ref{eq:rhoUCE})  not in the operator form, but via
the kernel of the photon-electron  matrix keeping temporarily the number of
components  fixed   (specifically,   $N$  photons  and  $\Ne$   electrons).
Simultaneously  we substitute  the time argument $t$ of the density  matrix
instead of $t_0$, because this argument is macroscopic time.

According to  equation~(\ref{eq:rhokernrho})  for photons and analogous for
electrons, we multiply equation (\ref{eq:rhoUCE}) by the corresponding state
vectors and their conjugates on the left and right.  We get
\beq \label{eq:rhos1sN}
&\!\!\!T_0\!\!\!& \DR{}{t}\rho_{s_1...s_N,\tau_1...\tau_{\Ne}}^{s'_1...s'_N,\tau'_1...
\tau'_{\Ne}}\biggl({\vkpr_1\atop\vk_1}{...\atop...}{\vkpr_N\atop\vk_N}
\biggl|{\vppr_1\atop\vp_1}{...\atop...}{\vppr_{{\Ne}}\atop\vp_{{\Ne}}}
\biggr|t\biggr)=\ovl{\Psi}_{s_1...s_N}(\vk_1,...,\vk_N)
\Phi^\dagger_{\tau_1...\tau_{\Ne}}(\vp_1,...,\vp_{\Ne}) \nonumber \\
&\times& \left[U^{\rm CE}_{\rm V}(t,t_0)\rho(t)-\rho(t)
U^{\rm CE}_{\rm V}(t,t_0)\right]\Psi_{s'_1...s'_N} (\vkpr_1,...,\vkpr_N)
\Phi_{\tau'_1...\tau'_{\Ne}}(\vppr_1,...,\vppr_{\Ne})N!{\Ne}! \nonumber \\
&=&i\frac{e^2}{c(2\pi\hbar)^9}\ovl{\Psi}_{s_1... s_N}
(\vk_1...\vk_N)\Phi^\dagger_{\tau_1... \tau_{\Ne}}(\vp_1... \vp_{\Ne})
\intl_\cV\d^4r'\intl_\cV\d^4r''\int\frac{\d^3k}{k}\frac{\d^3k'}{k'}
\frac{\d^3p}{p_0}\frac{\d^3p'}{p'_0} 
\cNatr46{s\ \tau\;}{s'\tau'}{\vk\;}{\vkpr}{\vp\;}{\vppr}{\ur''}{\ur'\;}
\nonumber \\
&\times& \int\frac{\d^3k''_1}{k''_1}\frac{\d^3k'''_1}{k'''_1}...
\frac{\d^3k''_N}{k''_N}\frac{\d^3k'''_N}{k'''_N}\frac{\d^3p''_1}{p''_{01}}
\frac{\d^3p'''_1}{p'''_{01}}...\frac{\d^3p''_{{\Ne}}}{p''_{0{\Ne}}}
\frac{\d^3p'''_{{\Ne}}}{p'''_{0{\Ne}}} 
\rho_{s''_1\,...s''_N,\tau''_1\;...\tau''_{\Ne}}^{s'''_1... s'''_N,
\tau'''_1...\tau'''_{\Ne}}\biggl({\vkpr''_1\atop\vkpr'_1}{...\atop...}
{\vkpr''_N\atop\vkpr'_N}\biggl|{\vp\,_1'''\atop\vppr'_1}{...\atop...}
{\vp\,_{\Ne}'''\atop\vppr'_{\Ne}}\biggr|t\biggr) \nonumber \\
&\times& \biggl[b_{\tau'}^{\dagger}
(\vppr)b_{\tau}(\vp)\ovl{a}_{(s')}(\vkpr)a_{(s)}(\vk)  
\Psi_{s''_1...s''_N} (\vkpr'_1,...,\vkpr'_N)
\Phi_{\tau''_1...\tau''_{\Ne}}(\vppr'_1,...,\vppr'_{\Ne})
\ovl{\Psi}_{s'''_1... s'''_N} (\vkpr''_1,...,\vkpr''_N)
\Phi^\dagger_{\tau'''_1...\tau'''_{\Ne}}(\vppr''_1,...,\vppr''_{\Ne}) \nonumber \\ 
&-& \Psi_{s''_1...s''_N}(\vkpr'_1,...,\vkpr'_N)
\Phi_{\tau''_1...\tau''_{\Ne}}(\vppr'_1,..., \vppr'_{\Ne})
\ovl{\Psi}_{s'''_1...s'''_N}(\vkpr''_1,...,\vkpr''_N)
\Phi^\dagger_{\tau'''_1...\tau'''_{\Ne}}(\vppr''_1,...,\vppr''_{\Ne}) \nonumber \\ 
&\times&b_{\tau}^{\dagger}(\vp)b_{\tau'}(\vppr)\ovl{a}_{(s)}(\vk){a}_{(s')}(\vkpr)
\biggr]{\Psi}_{s'_1... s'_N} (\vkpr_1,..., \vkpr_N)
\Phi_{\tau'_1... \tau'_{\Ne}}(\vppr_1,..., \vppr_{\Ne}).    
\eeq
Using    the    normalization     conditions     (\ref{eq:ortoPsi})     and
(\ref{eq:PhiPhiel})   for  the  state  vectors  as  well  as   formulae
(\ref{eq:aaPsi})  and  (\ref{eq:bdbPhi})  describing the
action of the creation and annihilation  operators on to the state vectors,
and  integrating  over all photon and electron  momenta  with two and three
primes, we get
\beq \label{eq:rhoN}
&\!\!\!T_0\!\!\!& \DR{}{t}\rho_{s_1...s_N,\tau_1...\tau_{\Ne}}
^{s'_1...s'_N,\tau'_1...\tau'_{\Ne}}
\biggl({\vkpr_1\atop\vk_1}{...\atop...}{\vkpr_N\atop\vk_N}
\biggl|{\vp\,_1'\atop\vp_1}{...\atop...}{\vp\,_{\Ne}'\atop\vp_{\Ne}}
\biggr|t\biggr)=i\frac{e^2}{c(2\pi\hbar)^9}\intl_\cV\d^4r'\intl_\cV\d^4r''
\int\frac{\d^3k}{k}\frac{\d^3k'}{k'}\frac{\d^3p}{p_0}\frac{\d^3p'}{p'_0}
\cNatr46{s\ \tau\;}{s'\tau'}{\vk\;}{\vkpr}{\vp\;}{\vppr}{\ur''}{\ur'\;} 
 \nonumber \\
&\times&
\sum_{r=1}^{N}\sum_{j=1}^{{\Ne}}\biggl[k'\delta_{s's_r}\delta(\vkpr-\vk_r)p'_0
\delta_{\tau'\tau_j}\delta(\vppr-\vp_j)  
\rho^{s'_1...s'_{r-1}s'_rs'_{r+1}...s'_N,\tau'_1 ... \tau'_{j-1}
\tau'_j\tau'_{j+1}...\tau'_{{\Ne}}}_{s_1...s_{r-1}s\ s_{r+1}... s_N,\tau_1...
\tau_{j-1}\tau\ \tau_{j+1}...\tau_{{\Ne}}}
\biggl({\vkpr_1\atop\vk_1}{...\atop...}{\vkpr_r\atop\vk}{...\atop...}
{\vkpr_N\atop\vk_N}\biggl|{\vppr_1\atop\vp_1}{...\atop...}{\vppr_j\atop\vp}
{...\atop...}{\vppr_{\Ne}\atop\vp_{\Ne}}\biggr|t\biggr) 
\nonumber \\
 &-& k\delta_{ss'_r}\delta(\vk-\vkpr_r)p_0
\delta_{\tau\tau'_j}\delta(\vp-\vppr_j)  
\rho^{s'_1...s'_{r-1}s'\,s'_{r+1}...s'_N,\tau'_1...\tau'_{j-1}\tau'
\tau'_{j+1}...\tau'_{{\Ne}}}_{s_1...s_{r-1}s_rs_{r+1}...s_N,\tau_1...\tau_{j-1}
\tau_j\tau_{j+1}...\tau_{{\Ne}}}\biggl({\vkpr_1\atop\vk_1}{...\atop...}
{\vkpr\atop\vk_r}{...\atop...}{\vkpr_N\atop\vk_N}\biggl|{\vppr_1\atop\vp_1}
{...\atop...}{\vppr\atop\vp_j}{...\atop...}{\vppr_{\Ne}\atop\vp_{\Ne}}
\biggr|t\biggr)\biggr]. 
\eeq
Deriving this equation, we used a number of times the possibility to change
the  summation   order  between  rows  and  columns  (see   derivation   of
eq.~[\ref{eq:ortoPsi}])  and as a result the factorials  canceled out.  The
difference  of two values of density  matrix in equation  (\ref{eq:rhoUCE})
gives the  difference  of  corresponding  matrix  elements  which we do not
write.

\subsection{Equations for the One-photon Matrix and the
Correlation Matrix}

Let us write now the equation for the kernel of the truncated photon matrix
of the first order which follows from equation  (\ref{eq:rhoN}).  We equate
all  corresponding  upper and lower  polarization  indices  and  momenta of
electrons and all (except the first)  indices and momenta of photons,  i.e.
we sum over indices and integrate over momenta.  The result,  according to the
definition (\ref{eq:rho1}), is then divided by $(N-1)!\Ne!$
\beq\label{eq:T0rho1}
&\!\!T_0\!\!&\DR{}{t}\rho_{s_1}^{s'_1}\left({\vk_1'\atop\vk_1}\biggl|t\right)= 
i\frac{e^2}{c(2\pi\hbar)^9}\intl_\cV\d^4r'\intl_\cV\d^4r''
\int\frac{\d^3k}{k}\frac{\d^3k'}{k'}\frac{\d^3p}{p_0}\frac{\d^3p'}{p'_0} 
\cNatr46{s\;\tau\;}{s'\tau'}{\vk}{\vkpr}{\vp}{\vppr}{\ur''}{\ur'}
\int\frac{\d^3k_2}{k_2}...\frac{\d^3k_N}{k_N}\frac{\d^3p_1}{p_{01}}...
\frac{\d^3p_{{\Ne}}}{p_{0{\Ne}}}  \nonumber \\
& \times& \frac{1}{(N\!-\!1)!{\Ne}!} 
\biggl[k'\delta_{s's_1}\delta(\vkpr-\vk_1)\sum_{j=1}^{{\Ne}}
p_0'\delta_{\tau'\tau_j}\delta(\vppr-\vp_j)\rho^{s'_1s_2... s_N,\tau_1\dots
\tau_j...\tau_{{\Ne}}}_{s\,\;s_2...s_N,\tau_1...\tau\,\;...\tau_{{\Ne}}}
\biggl({\vkpr_1\vk_2\atop\vk\,\;\vk_2}{...\atop...}{\vk_N\atop\vk_N}
\biggl|{\vp_1\atop\vp_1}{...\atop...}{\vp_j\atop\vp}
{...\atop...}{\vp_{{\Ne}}\atop\vp_{{\Ne}}} \biggr|t\biggr)  \nonumber \\
&+& \sum_{r=2}^Nk'\delta_{s's_r}\delta(\vkpr\!-\!\vk_r)\!
\sum_{j=1}^{{\Ne}}p_0'\delta_{\tau'\tau_j}\delta(\vppr\!-\!\vp_j)
\rho^{s'_1s_2...s_r...s_N,\tau_1...\tau_j...\tau_{{\Ne}}}_{s_1s_2...s\,\;...
s_N,\tau_1...\tau\,\;...\tau_{{\Ne}}}\biggl({\vkpr_1\vk_2\atop\vk\,\;\vk_2}
{...\atop...}{\vk_r\atop\vk}{...\atop...}{\vk_N\atop\vk_N}\biggl|
{\vp_1\atop\vp_1}{...\atop...}{\vp_j\atop\vp}{...\atop...}
{\vp_{\Ne}\atop\vp_{{\Ne}}}\biggr|t\biggr)  \nonumber \\
&-& k\delta_{ss'_1}\delta(\vk-\vkpr_1)\sum_{j=1}^{{\Ne}}
p_0\delta_{\tau\tau_j}\delta(\vp-\vppr_j)\rho^{s'\,s_2...s_N,\tau_1...\tau'...
\tau_{{\Ne}}}_{s_1s_2...s_N,\tau_1...\tau_j...\tau_{{\Ne}}}
\biggl({\vkpr\vk_2\atop\vk_1\vk_2}{...\atop...}{\vk_N\atop\vk_N}\biggl|
{\vp_1\atop\vp_1}{...\atop...}{\vppr\atop\vp_j}{...\atop...}{\vp_{{\Ne}}
\atop\vp_{{\Ne}}}\biggr|t\biggr) \nonumber \\
&-& \sum_{r=2}^N k\delta_{ss_r}\delta(\vk-\vk_r)\sum_{j=1}^{{\Ne}}
p_0\delta_{\tau\tau_j}\delta(\vp-\vp_j)\rho^{s'_1s_2...s'\,...s_N,\tau_1...
\tau'...\tau_{{\Ne}}}_{s_1s_2...s_r...s_N,\tau_1...\tau_j...\tau_{{\Ne}}}
\biggl({\vkpr_1\vk_2\atop\vk_1\vk_2}{...\atop...}{\vkpr\atop\vk_r}
{...\atop...}{\vk_N\atop\vk_N}\biggl|{\vp_1\atop\vp_1}{...\atop...}
{\vppr\atop\vp_j}{...\atop...}{\vp_{\Ne}\atop\vp_{{\Ne}}}\biggr|t\biggr) \biggr].
\eeq
Here in summation over $r$ we separately write the first term corresponding
to $r=1$.  Since every integral in the sums over $r$ and over $j$ gives the
same result, the summation is reduced to the  multiplication  by the number
of terms in the sums, so that the  factorials  cancel  out.  The sums  over
$r=2,...,N$    give   the   same   kernel   of   the    truncated    matrix
$\rho^{s'_1s'\!,\tau'}_{s_1s,\;\tau}$   and  cancel   out.  The   resulting
equation  in the  right  hand  side  contains  only  kernel  of the  matrix
depending on variables of one photon and one electron
\beq \label{eq:rhos1s1}
T_0\DR{}{t}\rho^{s'_1}_{s_1}\left({\vk_1'\atop\vk_1}\biggl|t\right) &=&
i\frac{e^2}{c(2\pi\hbar)^9}\intl_\cV\d^4r'\intl_\cV\d^4r''\int
\frac{\d^3k}{k}\frac{\d^3k'}{k'} \frac{\d^3p}{p_0}\frac{\d^3p'}{p'_0}
\cNatr46{s\ \tau\;}{s'\tau'}{\vk\;}{\vkpr}{\vp\;}{\vppr}{\ur''}{\ur'\;}  \nonumber\\
&\times& \left[k'\delta_{s's_1}\delta(\vkpr-\vk_1)\rho^{s'_1,\tau'}
_{s,\ \tau}\!\!\left({\vkpr_1\atop\vk\;}\biggl|{\vppr\atop\vp\;}\biggr|t\right)
-k\delta_{ss'_1}\delta(\vk-\vkpr_1)\rho^{s',\tau'}_{s_1,\tau}\!\!
\left({\vkpr\atop\vk_1}\biggl|{\vppr\atop\vp\;} \biggr| t \right)\right].  
\eeq

Let us now derive the equation for  $\rho^{s'_1,\tau'_1}_{s_1,\tau_1}$.  We
follow the same  procedure  as when  deriving  equation~(\ref{eq:rhos1s1}),
fixing the  characteristics  (momenta  and  indices)  of one photon and one
electron in equation  (\ref{eq:rhoN}).  One has to separate these variables
when we consider the action of annihilation and creation operators on them.
The  resulting  expression  contains  four couples of terms with pluses and
minuses.  The  first  couple  does  not  have   summation   at  all  (i.e.,
corresponds  to  $r=1,j=1$),  the  second  and the  third  couples  contain
summation  over the photon and electron  variables,  respectively,  and the
fourth  couple  contains  summations  over  variables  of both  interacting
particles  starting  from  $r=2$ and  $j=2$.  The terms in the last  couple
(with       double       sums)       give      the      same       function
$\rho^{s_1's'\!\!,\tau'_1\tau'}_{s_1s,\tau_1\tau}$ and annihilate.  We take
as the lhs of the equation the difference  which stands on the second place
of  equation  (\ref{eq:rhoUCE}).  We  omit  here  the  argument  $t$ of the
density matrices in the rhs of the equation based on the assumption  formulated
in the next section.  We get
\beq \label{eq:rhocor}
\rhomat{s_1',\tau_1'}{s_1,\tau_1}{\vkpr_1}{\vk_1}{\vppr_1}{\vp_1}{t} &-&
\rhomat{s_1',\tau_1'}{s_1,\tau_1}{\vkpr_1}{\vk_1}{\vppr_1}{\vp_1}{t_0}
 =i\frac{e^2}{c(2\pi\hbar)^9}\intl_\cV\d^4r'\intl_\cV\d^4r''
\int\frac{\d^3k}{k}\frac{\d^3k'}{k'}\frac{\d^3p}{p_0}\frac{\d^3p'}{p'_0}
\cNatr46{s\ \tau\;}{s'\tau'}{\vk\;}{\vkpr}{\vp\;}{\vppr}{\ur''}{\ur'\;}
\nonumber \\
&\times& \left[k'\delta_{s's_1}\delta(\vkpr-\vk_1)p_0'\delta_{\tau'\tau_1}
\delta(\vppr\!-\!\vp_1)
\rhoma{\rho}{s'_1,\tau'_1}{s\ ,\tau}{\vkpr_1}{\vk}{\vppr_1}{\vp}
- k \delta_{ss'_1}\delta(\vk-\vkpr_1) p_0\delta_{\tau\tau'_1}
\delta(\vp-\vppr_1)
\rhoma{\rho}{s'\,,\tau'}{s_1,\tau_1}{\vkpr}{\vk_1}{\vppr}{\vp_1} \right.\nonumber\\
&+& k'\delta_{s's_1}\delta(\vkpr-\vk_1)
\rhokpp{s_1',\tau'_1\tau'}{s\ ,\tau_1\tau}{\vkpr_1}{\vk}{\vppr_1}{\vp_1}
{\vppr}{\vp}-k\delta_{ss'_1}\delta(\vk-\vkpr_1)
\rhokpp{s'\,,\tau'_1\tau'}{s_1,\tau_1\tau}{\vkpr}{\vk_1}{\vppr_1}{\vp_1}
{\vppr}{\vp}  \nonumber \\
&+& \left. p_0'\delta_{\tau'\tau_1}\delta(\vppr-\vp_1)
\rhokkp{s'_1s'\!\!,\tau'_1}{s_1s,\tau}{\vkpr_1}{\vk_1}{\vkpr}{\vk}{\vppr_1}{\vp}
-p_0\delta_{\tau\tau_1'}\delta(\vp-\vppr_1)
\rhokkp{s'_1s'\!\!,\tau'}{s_1s,\tau_1}{\vkpr_1}{\vk_1}{\vkpr}{\vk}{\vppr}{\vp_1}
\right].   
\eeq
In order to proceed further we will use one additional assumption described in
the following section. Then we explore once more the smallness of the 
interaction scales in comparison with the scales of macroscopic changes.

\subsection{Approximation of Weakening of Correlations}

The matrices  depending on variables of a number of  particles  contain the
information about correlations between these particles.  In kinetic theory,
usually one applies an  approximation  of  molecular  chaos  (Silin  1971).
According  to that  approximation,  there are no  correlations  before  the
interaction, i.e.
\be \label{eq:nocor}
\rhomat{s'_1,\tau'_1}{s_1,\tau_1}{\vkpr_1}{\vk_1}{\vppr_1}{\vp_1}{t_0} =
\rhotwo{s'_1}{s_1}{\vkpr_1}{\vk_1} \rhotwo{\tau'_1}{\tau_1}{\vppr_1}{\vp_1} .
\ee
An  interaction   creates  the  correlations.  However,  according  to  the
principle of weakening of correlations  which is satisfied for sufficiently
rarefied  gases,  the  correlations  are  accounted  for  only in  equation
(\ref{eq:rhos1s1})  for the  one-particle  photon  matrix via kernel $\rho$
entering the right hand side of the  aforementioned  equation.  This kernel
is assumed to  characterize  the electron and photon states {\it after} the
interaction.  It can be  represented  via the same kernel {\it  before} the
interaction   and   the   correlation   function   as   it   is   done   in
equation~(\ref{eq:rhocor}).

The  two-particle  truncated  matrices can be presented  as products of the
one-particle   density   matrices   accounting  for  the  exchange  effects
(symmetrical  and   anti-symmetrical   forms  for  photons  and  electrons,
respectively)
\be \label{eq:induced}
\rhofour{s'_1s_2'}{s_1s_2}{\vkpr_1}{\vk_1}{\vkpr_2}{\vk_2}\!=\!
\rhotwo{s'_1}{s_1}{\vkpr_1}{\vk_1} \rhotwo{s'_2}{s_2}{\vkpr_2}{\vk_2}+
\rhotwo{s'_2}{s_1}{\vkpr_2}{\vk_1} \rhotwo{s'_1}{s_2}{\vkpr_1}{\vk_2}\!,
\  
\rhofour{\tau'_1\tau_2'}{\tau_1\tau_2}{\vppr_1}{\vp_1}{\vppr_2}{\vp_2}\!=\!
\rhotwo{\tau'_1}{\tau_1}{\vppr_1}{\vp_1} 
\rhotwo{\tau'_2}{\tau_2}{\vppr_2}{\vp_2}-
\rhotwo{\tau'_2}{\tau_1}{\vppr_2}{\vp_1} 
\rhotwo{\tau'_1}{\tau_2}{\vppr_1}{\vp_2}\!.
\ee
For matrices of higher order entering equation (\ref{eq:rhocor}), similar
equations hold
\be \label{eq:principle}
\rhokkp{s'_1s'_2,\tau'_1}{s_1s_2,\tau_1}{\vkpr_1}{\vk_1}{\vkpr_2}{\vk_2}
{\vppr_1}{\vp_1}=\rhofour{s'_1s_2'}{s_1s_2}{\vkpr_1}{\vk_1}{\vkpr_2}{\vk_2}
\rhotwo{\tau'_1}{\tau_1}{\vppr_1}{\vp_1}, \quad
\rhokpp{s'_1,\tau'_1\tau'_2}{s_1,\tau_1\tau_2}{\vkpr_1}{\vk_1}{\vppr_1}{\vp_1}
{\vppr_2}{\vp_2}=\rhotwo{s'_1}{s_1}{\vkpr_1}{\vk_1}
\rhofour{\tau'_1\tau_2'}{\tau_1\tau_2}{\vppr_1}{\vp_1}{\vppr_2}{\vp_2} .
\ee
It can be shown that these matrices satisfy all normalization conditions. 

Now we substitute equations (\ref{eq:nocor})--(\ref{eq:principle}) into the
rhs of  equation  (\ref{eq:rhocor}),  and  the  resulting  expression  into
equation  (\ref{eq:rhos1s1}).  At this step,  after all  substitutions,  we
will  use  the   fact   that   the   matrices   in  the  rhs  of   equation
(\ref{eq:rhocor})  do not depend on time  during the  interaction.  Then we
can take the integrals  over the  space-time  variables  enter the matrices
$\cN$ only.  Simultaneously we let $V\to\infty,\,  t_0\to-\infty$,  keeping
the upper limit $t$ finite.  As a result of this procedure  four  integrals
over   space   give  four   three-dimensional   $\delta$-functions   as  in
(\ref{eq:intNintM}), namely two pairs of the type
\be \label{eq:typesdelta1}
\delta(\vppr+\vkpr-\vp_{\rm v})\delta(\vp+\vk-\vp_{\rm v})=\delta(\vppr+\vkpr
-\vp-\vk)\delta(\vp+\vk-\vp_{\rm v})
\ee
and
\be \label{eq:typesdelta2}
\delta(\vppr-\vk-\vp_{\rm v})\delta(\vp-\vkpr-\vp_{\rm v})=\delta(\vppr+\vkpr
-\vp-\vk)\delta(\vp-\vkpr-\vp_{\rm v}).
\ee
Both  combinations  give the  $\delta$-functions  reflecting  the  momentum
conservation.  The second set of  $\delta$-functions gives a possibility to
take the integrals over $\vp_{\rm v}$.

Integrals over time with the upper limit $t$ can be taken independently. They
have the form
\be \label{eq:formintegr}
\intl_{-\infty}^te^{i(k'+p_0'-p_{\rm v}^0)ct'/\hbar}\d t'=
e^{i(k'+p_0'-p_{\rm v}^0)ct/\hbar}\left[\pi\hbar\delta(k'+p_0'-p_{\rm v}^0)
-\frac{i\hbar}{k'+p_0'-p_{\rm v}^0}\right].
\ee
Integrals  over  $p_{\rm  v}^0$ with the  second terms of this type must be
calculated  as the Cauchy principal  values.  These terms do not ensure the
energy  conservation  law.  They vanish if the limit of integration  $t$ is
equal to $\infty$.

The terms with absence of energy  conservation in kinetic equations, as was
noticed by Nagirner (1994), appear in Silin (1974) and Bomier (1991) who do
not discuss this fact.  In their papers, the interaction  between electrons
(Silin 1971) and between photons and atoms (Bomier 1991) does not depend on
time.  In our case, the  Hamiltonian  is more  complicated  and without the
energy  conservation law we cannot calculate the scattering  amplitudes and
the cross--sections of the process.

We suppose  that such  terms are not  physical  and must be  excluded.  The
interaction  between  photon and electron has  finished  before the density
matrix  changes  noticeably  and the  integration  limit  $t$ must be taken
infinite.   Then   the   resulting    integral   is   given   by   equation
(\ref{eq:intNintM}).  However, this procedure gives the cross-section twice
as large as the correct value.  To overcome this difficulty we continue the
upper limit $t$ to $\infty$ but take half of the result.

The next step is to use for the last time the assumption of the small scale
of the interaction.  Instead of the  one-particle  kernels we use now their
diagonal  in  momentum  forms   (\ref{eq:vigdia})   and   (\ref{eq:rhoss}).
Accounting   for   relation   (\ref{eq:deltak}),   the   lhs  of   equation
(\ref{eq:rhos1s1}) then takes the form
\be \label{eq:lhseq}
T_0\DR{}{t}\rho_{s_1}^{s'_1}\left({\vk_1'\atop\vk_1}\biggl|t\right)=
T_0k_1\delta(\vkpr_1-\vk_1)\frac{\d}{\d t}\rho^{s'_1}_{s_1}(\vk_1)=
k_1\delta(\uk'_1-\uk_1)\frac{2\pi\hbar}{c}\DR{}{t}\rho^{s'_1}_{s_1}(\vk_1).
\ee

After  all  the   substitutions   mentioned  above,  the  rhs  of  equation
(\ref{eq:rhos1s1})  contains 22 terms.  The two terms are the result of the
substitution   of   (\ref{eq:nocor}).  All   other   terms   in   equations
(\ref{eq:rhos1s1}) and (\ref{eq:rhocor}) come in couples, so that the total
number of terms is  multiple  of 4.  Every  term in the  first  line in the
square brackets of  (\ref{eq:rhocor})  gives two terms, and every term from
other two pairs of terms gives four terms, resulting in 20 terms.

The first two terms containing linearly the amplitude $M$ given by
equation (\ref{eq:amplMdef}) cancel out. This follows from the equality
\be \label{eq:Mdelta}
\matr44{s'\tau'}{s\;\tau\;}{\vk}{\vk}{\vp}{\vp}=
\delta_{ss'}\delta_{\tau\tau'} ,
\ee
which can be easily proven using  properties of the  solutions of the Dirac
equation. 
Equality  (\ref{eq:Mdelta})  leads  also to the  cancellation  of the eight
terms   arising   from   the   first   terms   in  the  rhs  of   equations
(\ref{eq:induced}).

All the remaining  twelve terms contain eight  three-dimensional  integrals
over photon and electron momenta.  There are two four-dimensional  and five
three-dimensional     $\delta$-functions     in    the    integrand.    All
three-dimensional  $\delta$-functions  disappear when taking the integrals.
The resulting  expression contains three  three-dimensional  integrals (one
over   photon   momenta   and  two   over   electron   momenta),   and  two
four-dimensional   $\delta$-functions,    $\delta(\up'+\uk'-\up-\uk)$   and
$\delta(\uk'_1-\uk_1)$.  The first one  reflects  the energy  and  momentum
conservation laws.  The second one cancels out since it also appears in the
lhs of equation (\ref{eq:rhos1s1}) according to  equation~(\ref{eq:lhseq}).

In the final  expression, four out of twelve terms contain  products of one
photon  and  one  electron   matrices   corresponding  to  the  spontaneous
scattering.  Two of these terms entering with the plus sign are responsible
for the  emission  and the two  terms  entering  with  the  minus  sign are
responsible for the  attenuation of the photon beam due to scattering.  The
remaining eight terms reflect the exchange  effects.  Four terms containing
the  product of one  electron  and two photon  matrices  correspond  to the
induced scattering, and four terms containing the product of one photon and
two  electron  matrices  reflect  the  exclusion   principle.  The  induced
scattering  terms have the same signs as the  spontaneous  terms, while the
exclusion principle terms have the opposite signs.  After an elementary but
very long deduction, we obtain the sought kinetic equation.

\section{Kinetic Equations for Photons and Electrons}

\subsection{Kinetic Equation for Polarized Photons}

As  we   mentioned   above,  all  the   remaining   terms  in  the  rhs  of
equation~(\ref{eq:rhos1s1}) after substitution of equation~(\ref{eq:nocor})
contain the factor  $\delta(\uk'_1-\uk_1)$  which also exists in the lhs of
the equation.  This factor  therefore  cancels out.  We can now restore the
dependence  of the photon  matrix on time and spatial  coordinates.  Let us
also change the  designations,  writing the photon matrix in the lhs in the
form  $\rho^{s'_0}_{s_0}(\vk)$.  Rewriting the derivative  over the line of
sight as the full  derivative,  the lhs of the equation takes the covariant
form
\be \label{eq:lhsform}
\uk \ \unb\rho^{s'_0}_{s_0}(\vk,\vr,t)=
\left( \frac{k}{c}\frac{\partial}{\partial t} +
\vk\cdot \vnb\right) \rho^{s'_0}_{s_0}(\vk,\vr,t).
\ee
In the rhs of the equation we change the indices and add and subtract two
pairs of identical terms containing products of two photon and two electron 
matrices for symmetry. The resulting kinetic equation takes the form 
\beq \label{eq:rkeph}
\uk \ \unb\rho^{s'_0}_{s_0}(\vk,\vr,t)&=& -\frac{r_0^2}{2}
\frac{m^2c^2}{(2\pi\hbar)^3}\int\frac{\d^3k'}{k'}\frac{\d^3p}{p_0}
\frac{\d^3p'}{p'_0}\delta(\up'+\uk'-\up-\uk)   \nonumber \\
&\times& \left\{\rho^{s''}_{s_0}(\vk)\rho^{\tau''}_{\tau'''}(\vp)
\matr44{s''\tau''}{s\;\;\tau}{\vk\;}{\vkpr}{\vp\;}{\vppr}
\left[\delta^{s'}_{s}+\rho^{s'}_{s}(\vkpr)\right]
\left[\delta^{\tau'}_{\tau}-\rho^{\tau'}_{\tau}(\vppr)\right]
\matr44{s'\,\tau'}{s_0'\tau'''}{\vkpr}{\vk\;}{\vppr}{\vp\;}  \right. \nonumber \\
&+&\matr44{s_0\tau'''}{s''\tau''}{\vk\;}{\vkpr}{\vp\;}{\vppr}
\left[\delta^{s'}_{s''}+\rho^{s'}_{s''}(\vkpr)\right]
\left[\delta^{\tau'}_{\tau''}-\rho^{\tau'}_{\tau''}(\vppr)\right]
\matr44{s'\tau'}{s\;\tau\,\;}{\vkpr}{\vk\;}{\vppr}{\vp\;}
\rho^{s'_0}_{s}(\vk) \rho^{\tau'''}_{\tau}(\vp) \nonumber \\
&-& \matr44{s_0\tau'''}{s''\tau''}{\vk\;}{\vkpr}{\vp\;}{\vppr}
\rho^{s'}_{s''}(\vkpr) \rho^{\tau'}_{\tau''}(\vppr)
\matr44{s'\tau'}{s\;\tau\;}{\vkpr}{\vk\;}{\vppr}{\vp\;}
\left[\delta^{s_0'}_{s}+\rho^{s'_0}_{s}(\vk)\right]
\left[\delta^{\tau'''}_{\tau}-\rho^{\tau'''}_{\tau}(\vp)\right]  \nonumber \\
&-& \left. \left[\delta^{s''}_{s_0}+\rho^{s''}_{s_0}(\vk)\right]
\left[\delta^{\tau''}_{\tau'''}-\rho^{\tau''}_{\tau'''}(\vp)\right]
\matr44{s''\tau''}{s\;\;\tau\,\;}{\vk\;}{\vkpr}{\vp\;}{\vppr}
\rho^{s'}_{s}(\vkpr)\rho^{\tau'}_{\tau}(\vppr)\matr44{s'\;\tau'}{s_0'\tau'''}
{\vkpr}{\vk\;}{\vppr}{\vp\;}\right\}.   
\eeq
Here  $r_0=e^2/mc^2$ is the classical electron radius.  The arguments $\vr$
and $t$ of the matrices in the  collision  integral are omitted.  The terms
in the first two lines in braces  describe  the photons  outgoing  from the
state  with  momentum  $\vk$,  the other two  lines  describe  the  photons
incoming  into this  state.  In order to outline  the matrix  character  of
Kronecker  symbol we use further the  notation  $\delta_s^{s'}$,  where the
lower index represents the rows and the upper index represents the columns.

Note that the products of four matrices are added just for the symmetry of
the equation. If these four terms are dropped and we change some indices, 
the equation   takes the form
\beq \label{eq:rkephot}
\uk\ \unb\rho^{s'_0}_{s_0}(\vk,\vr,t)&=& -\frac{r_0^2}{2}
\frac{m^2c^2}{(2\pi\hbar)^3}\int\frac{\d^3k'}{k'}\frac{\d^3p}{p_0}
\frac{\d^3p'}{p'_0}\delta(\up'+\uk'-\up-\uk)   \nonumber \\
&\times& \left\{\rho^{s''}_{s_0}(\vk)\rho^{\tau''}_{\tau'''}(\vp)\left[
M_{s''\tau''}^{s'\;\tau'}-M_{s''\tau''}^{s'\;\tau}\rho_\tau^{\tau'}(\vppr)+
M_{s''\tau''}^{s\;\;\tau'}\rho_s^{s'}(\vkpr)\right]M_{s'\;\tau'}^{s_0'\tau'''}
\right. \nonumber \\
&+& M_{s_0\,\tau'''}^{s''\tau''}\left[M_{s''\tau''}^{s\;\;\tau}-
\rho_{\tau''}^{\tau'}(\vppr)M_{s''\tau'}^{s\;\;\tau}+\rho_{s''}^{s'}(\vkpr)
M_{s'\tau''}^{s\;\tau}\right]\rho^{s'_0}_{s}(\vk) \rho^{\tau'''}_{\tau}(\vp) 
   \nonumber \\  
&-&2M_{s_0\tau''}^{s\;\,\tau}\rho_s^{s'}(\vkpr)\rho_\tau^{\tau'}(\vppr)
M_{s'\;\tau'}^{s_0'\tau''}+2M_{s_0\tau''}^{s\;\,\tau}\rho_s^{s'}(\vkpr)
\rho_\tau^{\tau'}(\vppr)M_{s'\;\tau'}^{s_0'\tau'''}
\rho_{\tau'''}^{\tau''}(\vp)   \nonumber \\  
&-& \left. M_{s_0\tau''}^{s\;\,\tau}\rho_s^{s'}(\vkpr)\rho_\tau^{\tau'}(\vppr)
M_{s'\;\tau'}^{s''\tau''}\rho_{s''}^{s_0'}(\vk)-\rho_{s_0}^{s''}(\vk)
M_{s''\tau''}^{s\;\;\tau}\rho_s^{s'}(\vkpr)\rho_\tau^{\tau'}(\vppr)
M_{s'\;\tau'}^{s_0'\tau''}\right\}.  
\eeq
In the last equation we omit the arguments of scattering amplitudes $M$ for
brevity.  These  arguments  are the same and in the same order as in equation
(\ref{eq:rkeph}).

We note, that a possibility to calculate the scattering  amplitudes appears
after we choose the  polarization  basis for photons and the axis where one
projects the electron spin.  As the axis one can use the electron  momenta.
The  situation is more  complex for the  polarization  bases.  We will give
detailed  description  of that later and now turn to the  equation  for the
electron distribution function.

\subsection{Kinetic Equation for Polarized Electrons}

Now we also can write down the kinetic  equation for  electrons.  It can be
deduced following the same procedure as in the case of the photon equation.
However,  due to the  symmetry of equation  (\ref{eq:rkeph}),  we can write
this  equation by analogy.  One just  removes one  integral  over  electron
momentum and adds one  integral  over photon  momentum as well as sums over
photon   polarization  and  removes  summation  over  indices  of  electron
polarization   (with  the   external   momentum,   $\vp$).  The   resulting
relativistic kinetic equation takes the form
\beq\label{eq:rkeel}
\up\ \unb\rho^{\tau'_0}_{\tau_0}(\vp,\vr,t)&=&
-\frac{r_0^2}{2} \frac{m^2c^2}{(2\pi\hbar)^3}\int\frac{\d^3k}{k}
\frac{\d^3k'}{k'} \frac{\d^3p'}{p'_0}\delta(\up'+\uk'-\up-\uk)  \nonumber \\
&\times& \left\{\rho^{s''}_{s'''}(\vk) \rho^{\tau''}_{\tau_0}(\vp)
\matr44{s''\tau''}{s\;\;\tau\,\;}{\vk\;}{\vkpr}{\vp\;}{\vppr}
\left[\delta^{s'}_{s}+\rho^{s'}_{s}(\vkpr)\right]
\left[\delta^{\tau'}_{\tau}-\rho^{\tau'}_{\tau}(\vppr)\right]
\matr44{s'\;\;\tau'}{s'''\tau'_0}{\vkpr}{\vk\;}{\vppr}{\vp\;} \right. \nonumber \\
&+& \matr44{s'''\tau_0}{s''\tau''}{\vk\;}{\vkpr}{\vp\;}{\vppr}
\left[\delta^{s'}_{s''}+\rho^{s'}_{s''}(\vkpr)\right]
\left[\delta^{\tau'}_{\tau''}-\rho^{\tau'}_{\tau''}(\vppr)\right]
\matr44{s'\tau'}{s\;\tau\;}{\vkpr}{\vk\;}{\vppr}{\vp\;}
\rho^{s_0'}_{s}(\vk) \rho^{\tau'_0}_{\tau}(\vp) \nonumber \\  
&-& \matr44{s'''\tau_0}{s''\tau''}{\vk\;}{\vkpr}{\vp\;}{\vppr}
\rho^{s'}_{s''}(\vkpr) \rho^{\tau'}_{\tau''}(\vppr)
\matr44{s'\tau'}{s\;\tau\;}{\vkpr}{\vk\;}{\vppr}{\vp\;}
\left[\delta^{s'''}_{s}+\rho^{s'''}_{s}(\vk)\right]
\left[\delta^{\tau'_0}_{\tau}-\rho^{\tau'_0}_{\tau}(\vp)\right]
 \nonumber \\  
&-& \left.  \left[\delta^{s''}_{s'''}+\rho^{s''}_{s'''}(\vk)\right]
\left[\delta^{\tau''}_{\tau'_0}-\rho^{\tau''}_{\tau'_0}(\vp)\right]
\matr44{s''\tau''}{s\;\;\tau\,\;}{\vk\;}{\vkpr}{\vp\;}{\vppr}
\rho^{s'}_{s}(\vkpr) \rho^{\tau'}_{\tau}(\vppr)
\matr44{s'\;\;\tau'}{s'''\tau'_0}{\vkpr}{\vk\;}{\vppr}{\vp\;} \right\}. 
\eeq

It is easy to check that in the arbitrary time moment, $t$, and in the
arbitrary point in space, $\vr$, two relations are valid
\be \label{eq:contin}
\unb\int\uk\;\rho^{s_0}_{s_0}(\vk,\vr,t)\frac{\d^3k}{k}=0, \quad
\unb\int\up\;\rho^{\tau_0}_{\tau_0}(\vp,\vr,t)\frac{\d^3p}{p_0}=0.
\ee
These  relations  reflect  the  continuity  equations  for the  photon  and
electron number and are the consequence of the  conservation  of the number
and quality of particles  during the scattering.  It is also easy to check
the validity of the energy-momentum conservation for Compton scattering
\be \label{eq:conserv}
\nabla_\mu\left[\int k^\mu k^\nu\rho^{s_0}_{s_0}(\vk,\vr,t)\frac{\d^3k}{k}+
\int p^\mu p^\nu\rho^{\tau_0}_{\tau_0}(\vp,\vr,t)\frac{\d^3p}{p_0}\right]=0.
\ee
We note that conservation laws (\ref{eq:contin}) and (\ref{eq:conserv})
are valid only for the occupation numbers, but not for the polarizations.

\subsection{Scattering Amplitudes}

Matrix elements (\ref{eq:amplMdef}) for Compton scattering (amplitudes) can
be rather easily calculated for a special choice of polarization  bases ---
internal  bases.  We take the unit vectors of these bases in the  following
form.  For the photon of momentum $\vk$
\be \label{eq:intort} 
\ue^{\rm in}_{(1)}(\vk)=\frac{\xi'\uk+\xi\uk'-q\up}{m\,c\,q\,\Delta},
\quad\ue^{\rm in}_{(2)}(\vk)=\frac{1}{m^3\,c^3\,q\,\Delta}
\left\{\vk\vkpr\vp,p_0\vk\times\vkpr+k\vkpr\times\vp+k'\vp\times\vk
\right\}.
\ee
Here the dimensionless scalar products
\be \label{eq:xixi1q}
\xi\equiv\uk \ \up/m^2c^2=\uk'\ \up'/m^2c^2, \quad
\xi'\equiv\uk'\ \up/m^2c^2=\uk\ \up'/m^2c^2, \quad
q\equiv\uk\ \uk'/m^2c^2,
\ee
and  $\Delta=\sqrt{2\xi\xi'/q-1}$.  The  double  definitions  of $\xi$  and
$\xi'$ in equation  (\ref{eq:xixi1q}) as well as the relation  $\xi=\xi'+q$
are the consequence of the conservation laws.

The values $\xi$ and $\xi'$ are the photon energies before and after
scattering in the frame where the electron before interaction is at rest.
If we denote the cosine of scattering angle in this frame as $\mu_0$ then
\be \label{eq:qmu0}
q=\xi\xi'(1-\mu_0), \quad \xi'=\frac{\xi}{1+\xi(1-\mu_0)}, \quad
\Delta=\sqrt{\frac{1+\mu_0}{1-\mu_0}}, \quad \mu_0=1+\frac{1}{\xi}-\frac{1}{\xi'}.
\ee
For the photon of momentum $\vkpr$, the unit vectors  are very similar
\be \label{eq:intortpr}
\ue^{\rm in}_{(1)}(\vkpr)=\ue^{\rm in}_{(1)}(\vk), \quad
\ue^{\rm in}_{(2)}(\vkpr)=-\ue^{\rm in}_{(2)}(\vk).
\ee
One  can  add a  vector  of the  photon  four-momentum  $\uk$  (or  $\uk'$)
multiplied by any real number to the  polarization  unit vectors.  By doing
so, one can make the time  components  of these unit vectors (as well as of
the external  vectors  $\ue(\vk)$)  equal to zero.  For such  vectors,  the
scattering amplitudes (matrix elements) which are marked with a circle atop
of the letter, are
\beq \label{eq:Msimple}
 \Mo{11}{11}&=&\Mo{12}{12}=\frac{1}{\Gamma}U_{+}\left(V_{+}-
\frac{2}{\Gamma}V_{-}\right), \qquad \  
\Mo{11}{12}=-\Mo{12}{11}=
\frac{1}{\Gamma}U_{-}\left(V_{-}-\frac{2}{\Gamma}V_{+}\right),  \nonumber \\
\Mo{11}{21}&=&-\Mo{12}{22}=i\frac{\xi}{\Gamma^3}(U_{-}V_{+}-U_{+}W_{+}),   \quad
\Mo{11}{22}=\Mo{12}{21}=i\frac{\xi}{\Gamma^3}(U_{+}V_{-}-U_{-}W_{-}),  \quad
\Mo{21}{21}=\Mo{22}{22}=\frac{1}{\Gamma}U_{+}V_{+},   \\
\Mo{21}{11}&=&-\Mo{22}{12}=i\frac{\xi'}{\Gamma^3}(U_{-}V_{+}+U_{+}W_{+}), \quad  
\Mo{21}{12}=\Mo{22}{11}=i\frac{\xi'}{\Gamma^3}(U_{+}V_{-}+U_{-}W_{-}), \quad
\Mo{21}{22}=-\Mo{22}{21}=\frac{1}{\Gamma}U_{-}V_{-}.
\nonumber
\eeq
Here we denoted
\beq \label{eq:UVW}
V_\pm&=&\sqrt{\Gamma+1}\sqrt{\Gamma'+1}\pm\sqrt{\Gamma-1}\sqrt{\Gamma'-1},  \quad
W_\pm=\sqrt{\Gamma+1}\sqrt{\Gamma'-1} \pm\sqrt{\Gamma-1}\sqrt{\Gamma'+1},\nonumber\\
U_\pm&=&\frac{1}{4}\frac{\Delta'+Z\pm\Delta}{\sqrt{\Delta'(\Delta'+ Z)}},\quad
\Gamma=\sqrt{\Delta^2+1}=\sqrt{\frac{2\xi\xi'}{q}}=\sqrt{\frac{2}{1-\mu_0}}, \\
\Gamma'&=&\frac{\Gamma}{2}\left(\frac{\xi}{\xi'}+\frac{\xi'}{\xi}\right)
=\frac{\Gamma^2+q}{\Gamma}, \quad 
Z=\frac{\Gamma}{2}q\left(\frac{1}{\xi'}+\frac{1}{\xi}\right)=
\frac{\xi+\xi'}{\Gamma},
\quad \Delta'=\sqrt{(\Gamma')^2-1}= \sqrt{\Delta^2+Z^2}. \nonumber
\eeq

Note that the scattering  amplitudes enter equations  (\ref{eq:rkeph})  and
(\ref{eq:rkeel})  without summation on their indices only with the products
of four  density  matrices  (two photon and two  electron  ones), but as we
mentioned these products  annihilate.  In all other terms, as we can see in
the  equation  for  photons  in  the  form  (\ref{eq:rkephot}),  there  are
summations over at least one pair of indices of scattering amplitudes.

If indeed the summation is over only one pair of indices then the resulting
sum  depends  on  the 6  remaining  indices  and  contains  $2^6=64$  terms
altogether.  If, on the other hand, the  summation  is  fulfilled  over two
pairs of indices,  then the sum  contains  $2^4=16$  terms.  The term which
describes the attenuation of radiation belongs to the last category
\be \label{eq:tMMdef}
t_{s'\tau'}^{s\;\tau}=\matr44{s'\;\tau'}{s''\tau''}{\vk\;}{\vkpr}{\vp\;}{\vppr}
\matr44{s''\tau''}{s\;\;\tau\;}{\vkpr}{\vk\;}{\vppr}{\vp\;} ,
\ee
where the sum is taken over one pair of photon indices and over one pair of
electron indices. Here one gets
\beq \label{eq:tMMres}
& \disp \tm{11}{12}=\tm{12}{11}=\tm{21}{22}=\tm{22}{21}=0, \quad \qquad \qquad
\qquad & \tm{11}{11}=\tm{12}{12}=\frac{B}{2}-1+\mu_0^2, \quad
\tm{21}{21}=\tm{22}{22}=\frac{B}{2}, \nonumber \\
& \disp \tm{11}{22}=\tm{12}{21}=-\tm{21}{12}=-\tm{22}{11}
=i\frac{q}{\Gamma^3}\left(1-\frac{1}{\xi'}\right), & 
\tm{11}{21}=\tm{22}{12}=-\tm{12}{22}=-\tm{21}{11}=
i\frac{\Delta}{\Gamma^3}(\xi'+\xi\mu_0),  
\eeq
where $B=\xi/\xi'+\xi'/\xi$.

\subsection{Transformation Matrices}

We  have  deduced  the  formulae  for  the  Compton  scattering  amplitudes
(\ref{eq:Msimple})  in  a  specially  chosen  internal  polarization  basis
(\ref{eq:intort}).  However,  the   expressions   (\ref{eq:Mmatrdef})   and
(\ref{eq:amplMdef})  for these  amplitudes  contain  unit  vectors  of some
external bases which should reflect the geometry of the scattering  medium.
The   photon   density    matrices    entering   the   kinetic    equations
(\ref{eq:rkeph})-(\ref{eq:rkeel})  are  connected  to the  same  (external)
vectors.  Therefore, we must find transformations  between the external and
the internal bases.

Two sets of polarization vectors $\ue_{(s)}(\vk)$ and $\ue_{(r)}^{\rm
\,\,in}(\vk)$ together with the vector $\vk/k$ form two three-dimensional
bases with one common unit vector. Consequently, the external vectors and the
internal vectors are connected to one another via relation
\be \label{eq:einlex}
\ue_{(s)}(\vk)=l^{.\,r}_{s\,.}(-\chi)\ue_{(r)}^{\rm \,\,in}(\vk),
\ee
where the angle $\chi$ is defined in the following manner
\be \label{eq:chidef}
\cos\chi=-\ue_{(1)}(\vk)\ue_{(1)}^{\rm \,\,in}(\vk)=
-\ue_{(2)}(\vk)\ue_{(2)}^{\rm \,\,in}(\vk),\quad
\sin\chi=\ue_{(1)}(\vk)\ue_{(2)}^{\rm \,\,in}(\vk)=
-\ue_{(2)}(\vk)\ue_{(1)}^{\rm \,\,in}(\vk).
\ee
The specific expressions for $\cos\chi$ and $\sin\chi$ are given in
Nagirner \& Poutanen (1993) where the external vectors 
are defined for the plane-parallel medium. The transformation matrix
\be \label{eq:lmatr}
\{l^{.\,r}_{s\,.}(\chi)\}=\left(\begin{array}{cc}
\cos\chi & \sin\chi \\ -\sin\chi & \cos\chi \\
\end{array}\right)
\ee
has the following properties
\be \label{eq:lprop}
l^{.\,r}_{s\,.}(\chi)=l_{.\,r}^{s\,.}(\chi)=l_{r\,.}^{.\,s}(-\chi)=
l^{r\,.}_{.\,s}(-\chi).
\ee
These properties are obvious if the numbers 1, 2 are substituted instead of
indices.  Matrices of this kind are  commutative  and  multiplicative,  and
their  transposition is equivalent to the inversion or changing the sign of
the argument.

Multiplying both sides of equations (\ref{eq:einlex}) by the Dirac matrices 
we get
\be \label{eq:eleDirac}
\he_{(s)}(\vk)=l^{.\,r}_{s\,.}(-\chi)\he^{\rm\,in}_{(r)}(\vk),\,\,
\he_{(s')}(\vkpr)=l^{.\,r'}_{s'\,.}(-\chi')\he^{\rm\,in}_{(r')}(\vkpr).
\ee
The  transformations  of bases related to the photon  momentum  $\vkpr$ are
analogous to (\ref{eq:einlex}) and the angle $\chi'$ is defined analogously
to   $\chi$   in   equation    (\ref{eq:chidef}).   Then,   for   amplitude
(\ref{eq:amplMdef}), we have
\beq \label{eq:MexMin}
 \matr44{s'\tau'}{s\;\tau\;}{\vkpr}{\vk\;}{\vppr}{\vp\;}
&=&\frac{1}{2}\ovl{u}^{\tau'}(\vppr)\left[\he_{(s)}(\vk)
\frac{mc+\hp-\hk'}{mc\xi'}\he_{(s')}(\vkpr)-\he_{(s')}(\vkpr)\frac{mc+\hp+\hk}
{mc\xi}\he_{(s)}(\vk)\right]u^\tau(\vp)   \nonumber \\
&=& l_{s\,.}^{.\,r}(-\chi)l_{s'\,.}^{.\,r'}(-\chi')\frac{1}{2}
\ovl{u}^{\tau'}(\vppr)\left[\he_{(r)}^{\rm\,in}(\vk)\frac{mc+\hp-\hk'}{mc\xi'}
\he_{(r')}^{\rm\,in}(\vkpr)-\he_{(r')}^{\rm\,in}(\vkpr)\frac{mc+\hp+\hk}{mc\xi}
\he_{(r)}^{\rm\,in}(\vk)\right]u^\tau(\vp)  \nonumber \\
&=& l_{s\,.}^{.\,r}(-\chi)l_{s'\,.}^{.\,r'}(-\chi')
\Motr{r'\tau'}{r\;\tau\;}{\vkpr}{\vk\;}{\vppr}{\vp\;} .   
\eeq
Using equations (\ref{eq:Mst}) and (\ref{eq:MexMin}) we obtain the 
following expression for the matrix element with the primed  upper  
momenta
\be \label{eq:MexMincnj}
 \matr44{s'\tau'}{s\;\tau\;}{\vk\;}{\vkpr}{\vp\;}{\vppr}
=\left[ \matr44{s\ \tau\;}{s'\tau'}{\vkpr}{\vk\;}{\vppr}{\vp\;} \right]^* 
=\left[ l^{s'.}_{.\,\,r'}(-\chi)l_{s\,.}^{.\,r}(-\chi')
\Motr{r\ \tau\;}{r'\tau'}{\vkpr}{\vk\;}{\vppr}{\vp\;} \right]^*   
= l^{s'.}_{.\,r'}(-\chi)l_{s\,.}^{.\,r}(-\chi')
\left[ \Motr{r'\tau'}{r\;\tau\;}{\vkpr}{\vk\;}{\vppr}{\vp\;} \right]^* \! .
\ee
These formulae must be substituted into equations (\ref{eq:rkeph}) and
(\ref{eq:rkeel}).

\subsection{Equation for Photons Interacting with Unpolarized Electrons}

If electrons are unpolarized (i.e., equation [\ref{eq:unpolele}] is valid),
then the matrix  $\rho^{\tau'}_{\tau}(\vp)$ must be replaced by the product
$\delta^{\tau'}_{\tau}(2\pi\hbar)^3f_\e(\vp)/2$.         The         factor
$\delta^{\tau'}_{\tau}-    \rho^{\tau'}_{\tau}(\vp)$    is    replaced   by
$\delta^{\tau'}_{\tau}[1-  (2\pi\hbar)^3f_\e(\vp)/2]$, and one can sum over
the electron polarizations.  For the external bases,
\be \label{eq:Tdef}
\tatr44{s's'''}{s\ s''}{\vkpr}{\vk\;}{\vppr}{\vp\;} 
=\matr44{s'\tau'}{s\ \tau}{\vk\;}{\vkpr}{\vp\;}{\vppr}
\matr44{s'''\tau}{s''\;\tau'}{\vkpr}{\vk\;}{\vppr}{\vp\;}=
\left[\matr44{s\ \tau\;}{s'\tau'}{\vkpr}{\vk\;}{\vppr}{\vp\;} \right]^*
\matr44{s'''\tau}{s''\;\tau'}{\vkpr}{\vk\;}{\vppr}{\vp\;} .
\ee
Here,  one  sums  over  the  repeated  indices  $\tau,   \tau'=1,2$.  After
summation over electron polarization the Compton scattering  cross-sections
of polarized  radiation  cannot be represented as a product  anymore.  This
reflects  the  transition  from the  ``pure''  scattering  of  photons  and
electrons with fixed polarization states to the mixed one.

The expressions for the cross--sections  (\ref{eq:Tdef})  have the simplest
form   in   the   internal   polarization   bases   (\ref{eq:intort})   and
(\ref{eq:intortpr}).   The   matrix    elements    are   then    given   by
(\ref{eq:Msimple}) and therefore
\be \label{eq:Todef}
\Totr{s's'''}{s\;s''}{\vkpr}{\vk\;}{\vppr}{\vp\;} =
\Motr{s'\tau'}{s\;\tau}{\vk\;}{\vkpr}{\vp\;}{\vppr}
\Motr{s'''\tau\;}{s''\,\tau'}{\vkpr}{\vk\;}{\vppr}{\vp\;} =
\left[\Motr{s\ \tau\;}{s'\tau'}{\vkpr}{\vk\;}{\vppr}{\vp\;}
\right]^*
\Motr{s'''\tau\;}{s''\,\tau'}{\vkpr}{\vk\;}{\vppr}{\vp\;}  ,
\ee
where the sums are taken over the repeated indices $\tau, \tau'=1,2$.

Only 8 out of 16 elements of matrix (\ref{eq:Todef}) are not zeros, and only 5
elements are different
\beq \label{eq:Tunpol}
\To{11}{11}&=&\frac{1}{2}[B+2+4(\mu_0^2-1)],\quad
\To{22}{22}=\frac{1}{2}(B+2), \quad
\To{21}{12}=\To{12}{21}=\frac{1}{2}(B-2),   \nonumber \\ 
\To{12}{12}&=&\To{21}{21}=\frac{1}{2}(B+2)\mu_0, \quad
\To{11}{22}=\To{22}{11}=-\frac{1}{2}(B-2)\mu_0.  
\eeq
The same expressions are obtained if they are derived directly for unpolarized
electrons by means of traces and projection operators 
(see e.g. Berestetskii et al. 1982).

Combination (\ref{eq:MexMin}) and (\ref{eq:MexMincnj}) with (\ref{eq:Tdef})
gives the law of transformation of the cross--sections of polarized radiation
by unpolarized electrons. One can write a transformation, for example, in the
following way
\be \label{eq:Ttrans}
\tatr44{s's'''}{s\ s''}{\vkpr}{\vk\;}{\vppr}{\vp\;} 
=l_{s'.}^{.\,r'}(-\chi)l^{s\,.}_{.\,r}(-\chi')
\Totr{r'r'''}{r\;r''}{\vkpr}{\vk\;}{\vppr}{\vp\;}
l^{.\,s''}_{r''\,.}(\chi)l_{.\,s'''}^{r'''\!.}(\chi').
\ee
The transformation matrices can be written in various forms using 
equation (\ref{eq:lprop}).

Keeping in mind the formulae expressing transformation laws and omitting the
arguments of the $T$-matrices, we rewrite the kinetic equation for photons 
with unpolarized electrons in the form
\beq 
 \uk\ \unb\rho^{s'_0}_{s_0}(\vk,\vr,t) &=& \frac{r_0^2}{2}
\frac{m^2c^2}{2}\int\frac{\d^3k'}{k'}\frac{\d^3p}{p_0}
\frac{\d^3p'}{p'_0}\delta(\up'+\uk'-\up-\uk)   \nonumber \\
& & \left\{ 
f_\e(\vppr) 
\left[1-(2\pi\hbar)^3f_\e(\vp)/2 \right]  
2T_{s_0s'}^{s\ s_0'}\rho_s^{s'}(\vkpr) - 
f_\e(\vp)  \left[1-(2\pi\hbar)^3f_\e(\vppr)/2 \right]  
\left[\rho_{s_0}^{s''}(\vk)T_{s''s'}^{s'\;s_0'} +  
T_{s_0\,s''}^{s''s} \rho_s^{s_0'}(\vk) \right] 
\right. \nonumber \\
& + & \left.  \left[ f_\e(\vppr) - f_\e(\vp)  \right] 
  \left[ \rho_{s_0}^{s''}(\vk) T_{s''s'}^{s\;\;s_0'}\rho_s^{s'}(\vkpr)  
+  T_{s_0\,s'}^{s''s}\rho_{s''}^{s'}(\vkpr) \rho_s^{s_0'}(\vk)   \right]
\right\}.  
\eeq
This equation can be easily rewritten in a more customary for astrophysics form
in terms of the Stokes parameters.

\subsection{Kinetic Equation in Terms of the Stokes Parameters}

The transformation from the polarization matrices to the Stokes parameters
is done with the aid of the formulae
\be \label{eq:rhoStokes}
\left\{\rho_s^{s'}\right\}=\left\{\begin{array}{cc}
n_{\rm I}+n_{\rm Q} & n_{\rm U}-in_{\rm V} \\
n_{\rm U}+in_{\rm V} & n_{\rm I}-n_{\rm Q} \\
\end{array}\right\},
\ee
where $\disp n_{\rm I}=  \rho_s^s(\vk)/2=\rho(\vk)$  is the mean occupation
number  of  photon  states,  $n_{\rm  Q},\,n_{\rm  U},\,n_{\rm  V}$ are the
corresponding   characteristics  of  photon   polarization   which  can  be
transformed  to  the  standard  Stokes  parameters   (having  dimension  of
intensity) by  multiplying  them by the factor  $2k^3c/(2\pi\hbar)^2$.  The
equation for polarized radiation interacting with unpolarized  electrons 
takes the following form in terms of the Stokes parameters 
\beq \label{eq:rkeStokes}
 \uk\ \unb\stn (\vk,\vr,t) & =& \frac{r_0^2}{2}m^2c^2\int
\frac{\d^3k'}{k'}\frac{\d^3p}{p_0}\frac{\d^3p'}{p_0'}
\delta(\up+\uk-\up'-\uk')\left\{f_\e(\vppr) [1-(2\pi\hbar)^3
f_\e(\vp)/2  ] \mL(-\chi)\mF\mL(\chi')\stn(\vkpr)  \right.  \nonumber \\
& +& \left. [f_\e(\vppr)-f_\e(\vp)]\mN(\vk) \mL(-\chi)\mF\mL(\chi')\stn(\vkpr)-
f_\e(\vp) [1-(2\pi\hbar)^3f_\e(\vppr)/2 ]
[F+\mL(-\chi)\mA\mL(\chi)]\stn(\vk) \right\},  
\eeq
where 
\be \label{eq:hatN}
\mN(\vk)= \left(\begin{array}{cccc}
n_{\rm I} & n_{\rm Q} & n_{\rm U} & n_{\rm V} \\
n_{\rm Q} & n_{\rm I} & 0 & 0 \\
n_{\rm U} & 0 & n_{\rm I} & 0 \\
n_{\rm V} & 0 & 0 & n_{\rm I} \\
\end{array}\right) , \quad  
\stn(\vk)= \left(\begin{array}{c}
n_{\rm I} \\
n_{\rm Q} \\
n_{\rm U} \\
n_{\rm V} \\
\end{array}\right) ,  
\quad  F=\mu_0^2-1+B,
\ee
\be \label{eq:hatFA}
\mF=\left(\begin{array}{cccc}
F & \mu_0^2-1 & 0 & 0 \\
\mu_0^2-1 &  \mu_0^2+1 & 0 & 0 \\
0 & 0 & 2\mu_0 & 0 \\
0 & 0 & 0 & B\mu_0 \\
\end{array}\right),
\quad  
\mA=\left(\begin{array}{cccc}
0 & \mu_0^2-1 & 0 & 0 \\
\mu_0^2-1 & 0 & 0 & 0 \\
0 & 0 & 0 & 0 \\
0 & 0 & 0 & 0 \\
\end{array}\right), 
\quad
\mL(\chi)=
\left(\begin{array}{cccc}
1 & 0 & 0 & 0 \\
0 & \ \cos 2\chi & \sin 2\chi & 0 \\
0 & -\sin 2\chi & \cos 2\chi & 0 \\
0 & 0 & 0 & 1 \\
\end{array}\right).
\ee
The   matrices   $\mL(\chi)$   have  the  same   properties   as   matrices
(\ref{eq:lmatr}).  They  are  commutative   and  their   transposition   is
equivalent  to the  inversion or changing  the sign of the  argument.  Note
that the same argument  $\chi$ appears with the opposite  signs in matrices
$\mL$   (around    matrix   $\mA$)   in   the   last   term   of   equation
(\ref{eq:rkeStokes}).  


As we have mentioned  above, the equation of the form  (\ref{eq:rkeStokes})
for non-degenerate  electrons (not accounting for the exclusion  principle)
was given in  Nagirner  (1994).  The term with  matrix  $\mA$  was  omitted
there.  It can be easily shown that this term  disappears  if the electrons
have an isotropic distribution (even if they are degenerate).  In the major
part of Nagirner's paper, the isotropic electron  distribution was assumed,
so that this omission did not introduced any errors.  \footnote{Let us note
here that the generalization of the Babuel-Peyrissac---Rouvillois  equation
to     polarized     radiation     (i.e.    a    limiting     version    of
eq.~[\ref{eq:rkeStokes}]  for weakly relativistic  Maxwellian electrons and
soft photons) was deduced in Nagirner  (1994).  The English  version of the
paper contains many misprints in the generalized  equation.  In the Russian
version,  $\tilde{n}$ should be substituted instead of $n$ in the first row
of his equation  (41) and the  brackets  before  $\tilde{n}''_1$  and after
$\tilde{n}_1'$  in the third row should be removed.  Hansen \& Lilje (1999)
who used the original equation of the form  (\ref{eq:rkeStokes})  presented
in Nagirner  (1994) also noted the  aforementioned  misprints.}  The matrix
$\mA$ was omitted also in papers of the authors (Nagirner \& Poutanen 1993,
1994) where the rhs of equation~(\ref{eq:rkeStokes})  was averaged over the
directions  of  electron  momenta  and the five  functions  describing  the
redistribution  of radiation in frequency,  angles and polarization  states
were  obtained.  However,  these  papers  were  devoted  to  the  isotropic
electrons as well.

Electron isotropy means that the medium is locally isotropic, and therefore
the attenuation (i.e.  the terms  corresponding to the last square brackets
in  eq.~[\ref{eq:rkeStokes}])  is described  not by the matrix, but only by
the  scalar  determined  by the  Klein---Nishina  cross-section,  $F$.  The
cross-section  averaged over momentum of scattered  photons can be found in
Nagirner \& Poutanen  (1994).  Hence we can  conclude  here that  polarized
electrons can influence both the linear and circular  polarization by means
of  attenuation  through   scattering.  Unpolarized,  even   non-isotropic,
electrons   can   introduce   and  change  in  this  process   only  linear
polarization.

\section{Conclusions}

In this paper, we have deduced the  relativistic  kinetic  equations  which
describe the behavior of the rarefied  photon and  electron  (or  positron)
gases  interacting  with each other via Compton  scattering.  We  accounted
here for stimulated effects for the photons and for the exclusion principle
for the electrons.  We considered arbitrary  polarization states of photons
and  electrons.  We  presented  also the  kinetic  equation  for  polarized
photons  scattered  by  unpolarized   electrons  in  terms  of  the  Stokes
parameters.   The   expressions   for   the   scattering   amplitudes   and
cross-sections  are derived  simultaneously.  There are no  limitations  on
photon and electron energies.

Note that all the  deductions  were made by means of  relativistic  quantum
electrodynamics methods and all the equations obtained are relativistically
covariant.  For  particular  scattering  problems  they, of course, must be
adapted to the geometry and symmetries of a medium and initial and boundary
conditions (see e.g. Nagirner \& Poutanen 1993).

Finally,  we would  like to  notice  that  the  factors  $1+\rho(\vk)$  and
$1-(2\pi\hbar)^3f_\e(\vp)/2$ accounting for the induced effects for photons
and the exclusion  principle for electrons in the scalar kinetic equations,
should be replaced by the new factors  $\delta_s^{s'}  +  \rho_s^{s'}(\vk)$
and $\delta_\tau^{\tau'} - \rho_\tau^{\tau'}(\vp)$ in the kinetic equations
for  polarized  photons  and  electrons.  This  rule can help to  formulate
kinetic equations for more complex processes for which a direct  derivation
could be very  complicated.  The  processes in strong  magnetic  fields can
serve as such examples.

\begin{acknowledgements}
This work was supported by the Swedish Natural  Science  Research  Council,
the  Anna-Greta  and Holger  Crafoord  Fund, the Royal Swedish  Academy of
Sciences,  the Russian  Federal  Program  ``Astronomia'',  and the  Russian
Leading  Scientific  Schools  grant  00-15-96607.  DIN is  grateful  to the
Stockholm Observatory for the hospitality during his visit.
\end{acknowledgements}

\end{document}